\documentstyle[11pt]{l-aa}
\input psfig
\begin{document}
\thesaurus{}
\title{Constraining $(\Omega,\Lambda)$ from weak lensing in clusters: the
triplet method.}
\author{L.Gautret$^{1}$, B. Fort$^{2}$, Y.Mellier$^{1,2}$}
\institute{$^{1}$DEMIRM, Observatoire de Paris, 61 Av. de 
l'Observatoire, 75014 Paris, France; and URA 336 du CNRS\\
$^{2}$Institut d'Astrophysique de Paris, 98 bis boulevard Arago,75014
Paris, France.}

%\thanks{<text of footnote>}
%\thanks{{\it Present address:\/} <address>}
\offprints{(Laurent Gautret) gautret@mesiom.obspm.fr}
\date{Received xx; accepted xx}
\maketitle

\begin{abstract}
We present a new geometrical method which uses weak gravitational lensing
effects around clusters to constrain  the cosmological parameters $\Omega$
and $\Lambda$ . \\
On each background galaxy, a cluster induces a {\sl convergence} and a {\sl shear} 
 terms, which depend on
the cluster projected potential, and on the
{\sl cosmological parameters} ($\Omega$, $\Lambda$),  through the angular distance
ratio $D_{LS}/D_{OS}$. To disentangle the effects of these three quantities,   
 we compare the relative values of the
measured ellipticities for each triplet of galaxies located at about the
same position in the lens plane, but having different color redshifts. The
simultaneous knowledge of the measured ellipticities and photometric
redshifts enable to build a purely geometrical estimator (hereafter the
$G(\Omega,\Lambda)$-estimator) independently of the  lens potential.
\\
More precisely $G$ has the simple form of the discriminant of a 3-3 matrix
built with the triplet values of $D_{LS}/D_{OS}$
and observed ellipticities. $G$ is then averaged on many triplets of
close-galaxies, giving a global function of ($\Omega$, $\Lambda$) which
converges to zero for the true values of the cosmological parameters.
\\
The linear form of $G$ regarding the measured ellipticity  of each galaxy
implies that  the different noises on $G$  decrease as $1/\sqrt{N}$, where
$N$ is the total number of observed distorted galaxies. A calculation and
comparison of each source of statistical noise is performed . \\
The possible systematics are analyzed with a multi-screen lensing model in
order 
 to estimate the  effect of perturbative potentials on galaxy triplets.
Improvements are then proposed to minimize these systematics and to
optimize the statistical signal to noise ratio. \\
Simulations are performed with realistic geometry and convergence for the 
lensing clusters and a redshift distribution for galaxies similar to what
is observed. They lead to the encouraging result that a significant
constraint on ($\Omega$, $\Lambda$) can be reached :
$\Lambda_{-0.2}^{+0.3}$ in the case $\Omega+\Lambda=1$ or
$\Omega_{-0.25}^{+0.3}$ in the case $\Lambda=0$ (at a 1$\sigma$ confidence
level). In particular the curvature of the Universe can be directly
constrained and the $\Omega=0.3$ and $\Omega=1$ universes can be separated
with a 2$\sigma$ confidence level. 
These constraints would be obtained from the observations of nearly 100
clusters. This corresponds to about 20 nights of VLT observations.The
method is still better adapted to a large program on the NGST.
  Hence, in complement to the supernovae method, the triplet method could in
principle clear up the issue of the existence and value of the cosmological
constant \\


\end{abstract}

\section{Introduction} 
Determination of the cosmological parameters of the standard cosmological
models is one of the great challenge for the next ten years.  Though these
are the main objectives of the MAP and Planck Surveyor satellites,
considerable efforts are devoted to the measurements of the ($\Omega$,
$\Lambda$)\footnote{$\Omega$ is the
matter density to critical density ratio. $\Lambda$ is the ratio between
the matter density associated
to the cosmological constant $\Lambda$ and the critical density. $\Lambda$
is equivalently defined as
the ratio $\Lambda/3H_{0}^{2}$, where $H_{o}$ is the Hubble constant.}
parameters prior to the launch of these
 surveyors.
 In this respect, the supernovae search (see the Supernova Cosmological
Project, Perlmutter \textit{et al.} 1998)
or gravitational lensing surveys (for a review see Mellier 1998) offer the
best perspectives.
  In particular, if they are used jointly, a  reasonable precision can be
reached, provided that the  degeneracy regarding to the determination of 
parameters $(\Omega,\Lambda)$ of each method are orthogonal. This 
is  for instance the case for the supernovae experiments and the weak
lensing analysis as presented hereinafter or as produced by large scale
structures (Van Waerbeke et al 1998).\\ 
The limitations of any  of these methods   
(even Planck measurements) are the understanding of the systematic bias and  
eventually the control of the large number of free parameters attached to
each of them. 
 Due to the difficulties to handle these issues it is important to diversify 
 the methods to measure $(\Omega , \Lambda)$ and to find out some new
observational tests.
 In this regard,  any new method that 
controls properly its own systematics and that decreases the number of
sensitive parameters needs careful attention.\\It is well known that
gravitational lensing can provide purely geometrical tests of the curvature
of the Universe. 
 Applications of this property to lensing clusters have been proposed by
Breimer \& Sanders (1992)  and Link \& Pierce (1998). They suggest to use
giant arcs having different redshifts to probe directly the curvature of
the Universe.  Evidently this method can provide the cosmological
parameters in the simplest way, provided that the modeling of the lens is
perfectly constrained. Besides, it requires the spectroscopic
redshifts of at least two different arcs in the same clusters, which is not   
 an easy task. As such, this is  a method which applies to very few
clusters. \\
Fort, Mellier \& Dantel-Fort (1996) 
 focused on a statistical approach which explores the  magnification bias 
 coupled with the redshift distribution of the sources.  Fort et al.  use
the shape and 
the extension of the depletion curves produced by the magnification of the
 galaxies
 to constrain the cosmological parameters and the redshift of the sources 
simultaneously. The intrinsic degeneracy can be somewhat broken if the
redshift of a giant arc is known and if the number density of 
 high-redshift  background galaxies is significant.
 However, in practice, reliable results need the investigation of a
significant number of lensing arc-clusters in order to improve the
statistics, to minimize the systematics (like multiple lens planes) and to 
 explore the sensitivity to the lens modeling.  \\
 The key issue on the Fort et al. approach is the coupling between the
cosmological
parameters, the redshift of the sources and the lens modeling. An attempt
to disentangle these
three quantities has been proposed by Lombardi \& Bertin  (1998), using a
method which 
 applies in rich clusters of galaxies, inside the region where the
weak-lensing regime is 
valid. They use 
the knowledge of the photometric redshifts for a joint  iterative
reconstruction of the 
mass of the cluster and the cosmology. However their iteration method
assumes that the mass
of the deflecting cluster is known
  (or equivalently they assume that the mass-sheet degeneracy inherent to
the mass reconstruction is  broken). This assumption is the key of the
problem, because  
 it means that  a perfect correction of the  systematic bias 
on the $(\Omega,\Lambda)$ determination due to the systematic errors  in 
the mass reconstruction (see section 3.1) can be achieved, which is still
not presently 
the case (see Mellier 1998 for a comprehensive review). 
 Indeed, for such a curvature test one should try to escape to the
uncertainties of the potential modeling (mass reconstruction)of the cluster. \\
The triplet method proposed in this paper can  solve this  problem
because 
 it is based on  the construction of an ($\Omega$,$\Lambda$)-estimator
independent of the lens potential.   
Basically, it consists in comparing the elliptical shear of 3 nearby galaxies
of different color redshifts probing the same part of the potential across
the cluster. When the galaxies are close enough, their observed 
ellipticity only depends on three unknown parameters,  the local
convergence and
shear (related to the second order derivatives of the projected potential
of the cluster)  and namely
the cosmological parameter ($\Omega, \Lambda$), through the angular
distance ratio 
$D_{LS}/D_{OS}$. The use of a triplet of nearby galaxies enables to break
this degeneracy and provides a local geometrical operator only dependent on
($\Omega, \Lambda$). We chose a local
 linear operator regarding to the observed ellipticities, and we 
  average it  on all set of close-triplets detected in many lens planes.
Follows a global geometrical estimator, biased by a noise (coming mainly
from the intrinsic source ellipticities) which decreases as the inverse of
the square root
of the number of triplets. It means that with a large  number of lensing
clusters one can estimate ($\Omega, \Lambda$) with a reasonable accuracy.
 This technique and its efficiency are discussed in the following sections.
 \\
 Section 2 reminds rapidly  the lensing equations and the basic lensing
quantities relevant for 
 the paper. Section 3 shows how to build the geometrical estimator $G$ 
  which uses  triplets of distorted  galaxies. The principle and the
detailed analysis 
of our method are also discussed.
 The signal to noise ratio of the method is then 
derived as well as the probability distribution of the cosmological
parameters 
($\Omega$,$\Lambda$).
 Although the main objective of this first paper is mostly to present the
principle of the method, Section 4 gives an evaluation of the amplitude of
several systematic biases coming from possible perturbing  lenses
distributed along the line of sight of triplets (galaxies or larger
structures). Preliminary solutions to these systematic biases and ideas of
optimizing the method (Selection in the geometry of clusters and choice of
redshifts) are developed in section 5.
 The  method is tested on simulations in Section 6.  Finally, we discuss the 
results and suggest some observational strategies in Section 7.\\

\section{The weak lensing equations}
The lensing properties are determined by the dimensionless convergence (the 
strength of the lens) $\kappa$ and shear (the distortion induced by the lens) 
\mbox{\boldmath $\gamma$}\footnote{In the following mathematical notations
with bold letters refer to 
complex numbers while usual letters are used for scalars or for the norm of
the 
associated complex numbers. The upper $^{*}$ index behind a complex number 
indicates its conjugated element.}, which both depend on the  second order
derivatives of the 
two-dimensional projected deflecting potential. The lensing effect of a
cluster on 
background galaxies can be expressed as an amplification matrix defined in
each angular position 
around the cluster as (Schneider et al. 1992)
%%%%%%%%%%%%%%%%%%%%%%%%%%%%%%%%%%
\begin{eqnarray}
    A^{-1}                      &=& \left( \begin{array}{lr}
                 1-\kappa-\gamma_{1} & -\gamma_{2}                \\
                 -\gamma_{2}                & 1-\kappa+\gamma_{1} 
                                           \end{array}
                                                           \right)  \ ,  
\end{eqnarray}
\noindent where
\begin{equation}
    \mbox{\boldmath $\gamma$} = \gamma_{1} + i\gamma_{2} .              
\end{equation}
%%%%%%%%%%%%%%%%%%%%%%%%%%%%%
\noindent $\kappa$ is a dimensionless form of the cluster surface mass
density $\Sigma$ :
%%%%%%%%%%%%%%%%%%%%%%%%%%%%%%%%%%%
\begin{eqnarray}
 \kappa          =  \Sigma/\Sigma_{crit} \ \mbox{with} \
 \Sigma_{crit} =\frac{c^2}{4 \pi G D_{OL}}\frac{D_{OS}}{D_{LS}}  \ ,
\end{eqnarray}
%%%%%%%%%%%%%%%%%%%%%%%%%%
\noindent where $D_{OS}$, $D_{OL}$ and $D_{LS}$ are the angular diameter
distances 
from the observer to the source, from the observer to the lens and from the
lens to the 
source respectively.  For the weak lensing regime the gravitational
distortion produced by a lensing cluster can be modeled by a transformation
in the 
ellipticity of the galaxies from the source plan ($\epsilon_{S}$) to the
image plane 
($\epsilon$) (see Appendix A) :
%%%%%%%%%%%%%%%%%%%%%%%%%%%%%
\begin{eqnarray}
  \mbox{\boldmath $\epsilon$} = (1-g^{2})\mbox{\boldmath $\epsilon$}_{S} + 
\mbox{\boldmath $g$} = \bar{\mbox{\boldmath $\epsilon$}}_{S} + 
\mbox{\boldmath $g$} \ \mbox{\ \ with\ \ } \ \mbox{\boldmath
$g$}=\frac{\mbox{\boldmath 
$\gamma$}}{1-\kappa}.
\end{eqnarray}
%%%%%%%%%%%%%%%%%%%%%%%%%%%%
$\mbox{\boldmath $\epsilon$}$ is the complex observed ellipticity,
$\mbox{\boldmath $g$}$ is the 
 complex reduced shear and $\bar{\mbox{\boldmath $\epsilon$}}_{S}$ is what
we will call the 
complex corrected source ellipticity.
Either in the source or the image plan, the ellipticity parameter is
defined by :
%%%%%%%%%%%%%%%%%%%%%%%%%%%%
\begin{eqnarray}
   \mbox{\boldmath $\epsilon$} =  \epsilon \ e^{2i\theta} \ \mbox{with} \
    \epsilon = \frac{1-r}{1+r} ;
\end{eqnarray}
%%%%%%%%%%%%%%%%%%%%%%%%%%%%%%%
\noindent $r$ is the axis ratio of the image isophotes and $\theta$ is the
orientation of the main 
axis.\\
The convergence and the shear both depend on the source redshift through an 
absolute lensing factor $\omega_{a}$ appearing in equation (3) :
%%%%%%%%%%%%%%%%%%%%%%%%%%%%%%%%
\begin{eqnarray}
  \omega_{a}(z)=\frac{D_{LS}}{D_{OS}} \ .
\end{eqnarray}

Adopting the notation of Seitz \& Schneider  (1997) who relate the lensing 
parameters to the value they would have at infinite redshift, $\kappa$ and 
$\gamma$ now write :
%%%%%%%%%%%%%%%%%%%%%%%
\begin{eqnarray}
  \kappa   &=&  \omega(z) \kappa_{\infty} \ , \\
  \mbox{\boldmath $\gamma$}  &=&  \omega(z) \mbox{\boldmath 
$\gamma$}_{\infty}  \ , \ {\rm where} \\
  \omega(z)  &=& \frac{\omega_{a}(z)}{\omega_{a}(\infty)} \ .
\end{eqnarray}

Hereafter, $\omega(z)$  will be named the lensing factor.  This is 
 the term which contains the cosmological dependency. 

\section{The method}
The behavior of $\omega(z)$  with $\Omega$ in the case $\Omega+\Lambda=1$ 
(flat geometry) and in the case $\Lambda=0$,  is given on figure (1). All
curves 
range from $0$ (for a source redshift equal to the cluster redshift) to $1$
(for an 
 infinite  source redshift). Their main difference is a small change in
their convexity. That is why the use of numerous triplets of sources at
different redshifts may disentangle these curves, i.e. provide a constraint on 
the cosmological parameters. \\
%%%%%%%%%%%%%%%%%%%%%%%%%%
\begin{figure}
\begin{center}
\psfig{figure=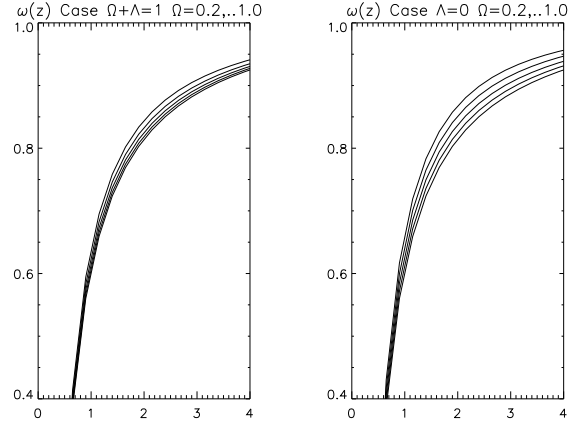,height=6cm}
\caption{$\omega$ versus redshift in the cases $\Omega+\Lambda=1$ and 
$\Lambda=0$, for a cluster lying at a redshift $0.4$. Different curves
correspond
to $\Omega=$ $0.2$ (bottom curve), $0.4,0.6,0.8$ and $1.0$  (top curve).
All curves 
start from $0$ (at redshift $0.4$) and converge to $1.0$ when the redshift
becomes infinite. 
The triplet method applies to their different relative convexity. }
\end{center}
\end{figure}
%%%%%%%%%%%%%%%%%%%%%%%%%%
The core of the method is to proceed in such a way that 
this constraint is independent on the potential of the cluster.
 We have  constructed  an operator which depends theoretically 
  on ($\Omega,\Lambda$) and can be computed simply from the observed 
 ellipticities of background galaxies as well as their photometric redshift.
 Its main  property is to be equal to zero when the cosmological parameters
are 
equal to the actual ones. We proceed in two steps: 
first, we build such an operator ($G_{ijk}$) from triplets of close
sources, and second we 
average it on many triplets of sources to obtain the final geometrical operator $G$.
%%%%%%%%%%%%%%%%%%%%%%%%%%%%%%%
\subsection{Construction of $G_{ijk}$}
For a background galaxy at redshift $z_{i}$, equation (4) rewrites
%%%%%%%%%%%%%%%%%%%%%%
\begin{eqnarray}
   \mbox{\boldmath $\epsilon$}_{i} &=& \bar{\mbox{\boldmath
$\epsilon$}}_{S,i} + 
\mbox{\boldmath $g$}_{i}^{o} \ \mbox{with} \\
   \mbox{\boldmath $g$}_{i}^{o} &=& \frac{\omega_{i}^{o} \mbox{\boldmath 
$\gamma$}_{\infty}}{1-\omega_{i}^{o} \kappa_{\infty}} ,
\end{eqnarray}
%%%%%%%%%%%%%%%%% 
\noindent where the lower index ($i$) refers to the redshift $z_{i}$ and the 
upper  index ($^{o}$)
refers to the actual values of the cosmological parameters : 
$\Omega_{o}$,$\Lambda_{o}$. The second term in equation (10),
$\mbox{\boldmath $g$}_{i}^{o}$ 
represents  the 
part of the image ellipticity that depends on the  cosmology.\\
Let us now consider a triplet of background neighboring galaxies in the
image plane
and lying at redshifts $z_{i}$, $z_{j}$ and $z_{k}$. The number density of
triplets 
depends on the deepness of the observations. For instance, up to $R=26$
mag.,  
 we expect a mean density of $50$ sources arcmin$^{-2}$, that is  about 
 4 sources inside  a circle of radius $10$ arcseconds. 
In the following, we thus consider that each galaxy of the triplet is
distorted by the same 
local potential i.e. $\kappa_{\infty}$ and $\mbox{\boldmath
$\gamma$}_{\infty}$ are 
the same for the three galaxies. The bias induced by this approximation
will be  
 discussed in section 3.4.
The triplet of galaxies gives a triplet of equations (11) respectively 
indexed by $i$, $j$ and $k$ from which we can derive a final equation
independent 
both on $\kappa_{\infty}$ and $\mbox{\boldmath $\gamma$}_{\infty}$. This 
equation writes simply as the zero of a 3-3 determinant :
%%%%%%%%%%%%%%%%%%%%%%%%
\begin{eqnarray}
   \left| \begin{array}{lcr}
      1 & \omega_{i}^{o} & \omega_{i}^{o}g_{j}^{o}g_{k}^{o} \\
      1 & \omega_{j}^{o} & \omega_{j}^{o}g_{k}^{o}g_{i}^{o} \\ 
      1 & \omega_{k}^{o} & \omega_{k}^{o}g_{i}^{o}g_{j}^{o} \\
              \end{array}  \right|    =  0   \  . 
\end{eqnarray}
%%%%%%%%%%%%%%%%%%%%%%%%%%%
The first term of this equation can now be formally generalized to a complex 
operator $G_{ijk}$ of 
$(\Omega,\Lambda)$ built  from the complex measured ellipticities of the
three 
galaxies :
%%%%%%%%%%%%%%%%%%%%%%%%%%%%%%%
\begin{eqnarray}
    G_{ijk}(\Omega,\Lambda) =  \left| \begin{array}{lcr}
      1 & \omega_{i} & \omega_{i} \mbox{\boldmath $\epsilon$}_{j}
                       
\mbox{\boldmath $\epsilon$}_{k}^{*} \\
      1 & \omega_{j} & \omega_{j}  \mbox{\boldmath $\epsilon$}_{k}
                        
\mbox{\boldmath $\epsilon$}_{i}^{*}\\ 
      1 & \omega_{k} & \omega_{k} \mbox{\boldmath $\epsilon$}_{i}
                       
\mbox{\boldmath $\epsilon$}_{j}^{*} \\
                 \end{array}  \right|      \ . 
\end{eqnarray}
%%%%%%%%%%%%%%%%%%%%%%%%%%%
The dependency in $(\Omega,\Lambda)$ is contained in each term
$\omega_{s}(\Omega,\Lambda)$
 ($s=i$, $j$ or $k$) defined by equation (9) for $z=z_{s}$. 
$G_{ijk}$ is more explicitly the sum of two $(\Omega,\Lambda)$ functions: 
$G_{ijk}^{main}$ which is equal to zero for the actual values of the
cosmological 
parameters, and a complex noise $N$:
\begin{equation}
  G_{ijk} = G_{ijk}^{main} + N  \ ,
\end{equation}
\noindent where
\begin{eqnarray}
    G_{ijk}^{main} &=&  \left| \begin{array}{lcr}
      1 & \omega_{i} & \omega_{i}/\omega_{i}^{o} \\
      1 & \omega_{j} & \omega_{j}/\omega_{j}^{o} \\ 
      1 & \omega_{k} & \omega_{k}/\omega_{k}^{o}\\
                 \end{array}  \right|
\omega_{i}^{o}\omega_{j}^{o}\omega_{k}^{o}\ 
\gamma_{\infty}^{2}    . 
\end{eqnarray}
%%%%%%%%%%%%%%%%%%%%%%%%%%%%%%%
At this point it is important to stress that the $\kappa_{\infty}$
contribution cannot be neglected (see equation (11)). Indeed, if we
consider for instance a  $0.1$ variation of the cosmological
parameters (along the gradient of $\omega(\Omega,\Lambda)$), the resulting
relative variation
of the term $\omega\gamma$ is about $1\%$ which is about ten times smaller
than the relative variation due to the $1-\omega\kappa$ term, in equation
(11). 
That is why both contributions from the local convergence
($\kappa_{\infty}$) and the local shear ($\gamma_{\infty}$) of the cluster
potential must be taken in account.\\
 Lombardi \& Bertin (1998) proposed
to reconstruct jointly the shear, the convergence and $(\Omega,\Lambda)$ in
the weak lensing
area, with an iteration method based on the equation (11). Their method
seems to converge relatively rapidly with a small number of clusters but it
seems that for their simulations they implicitly assume that the mass of
the cluster is known. However, one can see that     
equation (11) is invariant when replacing
$\gamma_{\infty}$ by $\alpha\gamma_{\infty}$ and $1-\omega\kappa_{\infty}$
by $1-\alpha\omega\kappa_{\infty}$ ($\alpha$ is a constant). This expresses
the 
mass-sheet degeneracy problem which implies that the total mass of
the lensing 
cluster is uncertain. Indeed, despite the numerous suggestions which have
been proposed in order 
 to solve this issue (Seitz \& Schneider, 1997), for the moment 
mass reconstruction
techniques cannot disregard systematic errors analysis (see Mellier 1998 for a 
comprehensive review). As an 
  example, a $20\%$ systematic bias on the  determination
of the total mass of the cluster (or equivalently a $20\%$ systematic on
the mean value of $\kappa_{\infty}$) is equivalent to a systematic bias
larger than 0.2 on the value of the
cosmological parameters (when compared to the 
$1\%$ contribution from the lensing factor mentioned above).  
Therefore, the knowledge of the lens potential is a critical 
 strong assumption.
\\
In the triplet method it is possible to constrain the cosmological
parameters regardless the  potential of the lens. No assumption is made in
order to relate
the values of the local shear and the local convergence. Hence they are
considered as independent parameters. In order to construct a $G$ operator
which only depends on $(\Omega,\Lambda)$, 
we then need three local equations relating $\kappa_{\infty}$,
$\gamma_{\infty}$ and $(\Omega,\Lambda)$ to
cancel the potential dependency. This is achieved with the measured
ellipticity equations (10) applied to triplets of
close galaxies at different redshifts. Besides, the form of the $G$
operator becomes unique and 
 must have the formal expression given in
equation (13) if we want it linear with respect to the ellipticities
provided by the
observations.
To sum up, the use of triplets of galaxies through the operator $G$ is the 
 simplest  way to build a pure 
geometrical operator which drops both the $\kappa_{\infty}$ and 
$\gamma_{\infty}$ dependencies and keeps linear regarding to the
ellipticities.\\

The statistical noise $N$ will be more explicitly calculated in section
3.4. For the moment it is just important to understand that the probability
distribution of this noise (real and imaginary parts) regarding the
different triplets of galaxies is a random law centered on 0  since the
linear construction of $G$ makes the different sources of noise  (mainly
the intrinsic source ellipticity) be  randomly distributed around 0.\\
The above formula does not take in account the systematics (an effect of
galaxy-galaxy lensing, a presence of background structures) that will be
studied further in section 4. 

%%%%%%%%%%%%%%%%%%%%%%%%%%%%%%%%%
\subsection{Construction of $G$}
Before averaging $G_{ijk}(\Omega,\Lambda)$ on many triplets (that is to say
on all 
the triplets done with galaxies contained within a given radius), let us
stress that it makes sense. Indeed, by averaging  (we only 
consider the real part of the complex noise) on many triplets  the noise
contribution 
 decreases as  
$1/\sqrt{N}$, with $N$ the number of background galaxies (and not the
number of 
triplets, since many triplets are redundant). While the noise is vanishing,
the method 
can provide valuable constraints as far as cosmological operator
$G_{ijk}^{main}$ does not 
vanish too,  
or in other words if its behavior with $\Omega$ and $\Lambda$ is the same
(not 
random) whatever the triplet is. Fortunately this is the case provided that
the three 
redshifts of all the triplets are ordered similarly : for example $z_{i} <
z_{j} < z_{k}$. 
\\
Consequently we can derive $G$ from an average of $G_{ijk}$ on all the
ordered 
triplets ${ijk}$ :
%%%%%%%%%%%%%%%%%%%%%%%%%%%%%
\begin{eqnarray}
   G(\Omega,\Lambda) = <G_{ijk}>_{(i<j<k)} \ .
\end{eqnarray}
%%%%%%%%%%%%%%%%%
According to equation (14), $G$ is also composed of two terms : the first one 
$G^{main}$ which has the same properties as $G_{ijk}^{main}$, and a
Gaussian noise $GN$ 
 which decreases as $1/\sqrt{N}$,
\begin{equation}
   R_e(G)(\Omega,\Lambda) = G^{main}(\Omega,\Lambda) + Re(GN) \ ,
\end{equation}
\noindent where
\begin{eqnarray}   
   G^{main} =  \left <\left| \begin{array}{lcr}
      1 & \omega_{i} & \omega_{i}/\omega_{i}^{o} \\
      1 & \omega_{j} & \omega_{j}/\omega_{j}^{o} \\ 
      1 & \omega_{k} & \omega_{k}/\omega_{k}^{o}\\
                 \end{array}  
\right|\omega_{i}^{o}\omega_{j}^{o}\omega_{k}^{o}\right >_{(ijk)}  
\left <\gamma_{\infty}^{2} \right >  \ ,
\end{eqnarray}
\noindent and
\begin{equation}
R_e(GN) \propto 1/\sqrt{N}  \ ,
\end{equation}
\noindent where $R_e$ denotes the real part of the complex quantities. 
$<\gamma_{\infty}^{2}>$ is the square of $\gamma_{\infty}$ averaged on the
weak 
lensing area. We consider the real part of $G$ ($Re(G)$) because as $N$,
the noise 
$GN$ is complex.\\
By construction $G^{main}$ is equal to zero at the position ($\Omega_{o}$, 
$\Lambda_{o}$). We thus can write the following equation (biased by
the presence 
of the Gaussian noise) which sums up the method : ($\Omega_{o}$,
$\Lambda_{o}$) is a solution of 
%%%%%%%%%%%%%%%%%%%%
\begin{eqnarray}
   G(\Omega,\Lambda) \equiv 0 .
\end{eqnarray}
%%%%%%%%%%%%%%%%%%%
In order to see if the triplet method can be effective from the
observational point of view 
two questions have to be addressed: 
\begin{enumerate}
 \item how is the $G$-operator degenerated in $(\Omega,\Lambda)$?
 \item how many clusters are necessary to cancel the noise contribution with
 respect to the cosmological information contained in $G^{main}$ ? 
\end{enumerate}
%%%%%%%%%%%%%%%%%%%%%%%%%%
\subsection{The main cosmological term: $G^{main}(\Omega , \Lambda)$}
Figure (2) gives the contours of $G^{main}/<\gamma_{\infty}^{2}>$ in the 
$(\Omega,\Lambda)$  plan, for the redshift distribution (42), as taken in the 
simulations (see section 6). From this graph we  see that the method is 
degenerated in $(\Omega,\Lambda)$. The degeneracy is parallel to the
$G$-contours. To break this degeneracy one has either to 
make a theoretical assumption (for example $\Omega + \Lambda=1$ ) or to add 
another experimental degenerated constraint which contours are as
orthogonal to the 
$G$-contours as possible, like for example 
 high-redshift supernovae (Perlmutter et al. 1998).
Considering the mean orientation of its degeneracy, the triplet method can
also be seen as a direct measure of the curvature of the Universe
$1-\Omega-\Lambda$.\\
From figure (2) we also get quantitatively the variations of $G$. It shows
that,   
for an accuracy of about $10\%$ on the cosmological parameters (along the
gradient of $G$), we must obtain a precision  of
about $10^{-4} <\gamma_{\infty}^{2}>$ on G. 
This rate of variation has to be compared with the noise on $G$.
%%%%%%%%%%%%%%%%%%%%%%
\begin{figure}
\begin{center}
\psfig{figure=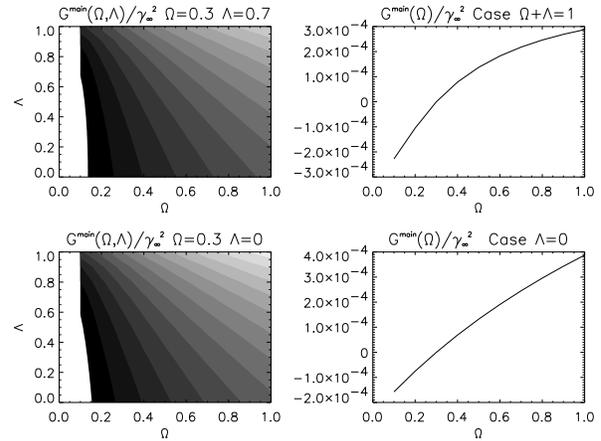,height=6cm}
\caption{Contours of $G^{main}(\Omega,\Lambda)/<\gamma_{\infty}^{2}>$ in 
the cases $(\Omega,\Lambda)=(0.3,0.7)$ (top panel) and $(0.3,0)$ (bottom
panel). 
The right part of the figure gives the restriction of $G^{main}$ to the
particular 
domains $\Omega+\Lambda=1$ (top) and $\Lambda=0$ (bottom).}
\end{center}
\end{figure}
%%%%%%%%%%%%%%%%%%%%%%
\subsection{Noise on G}
The complex noise is produced by four sources:
\begin{enumerate}
 \item noise from the corrected source ellipticities
$\bar{\epsilon}_{S,i,j,k}$,
 \item  the errors propagation on the measured ellipticities
$\Delta\epsilon_{i,j,k}$ (it behaves similarly as the previous ones),
 \item the three sources do not have the same
$\gamma_{\infty}$ and we thus have to consider
$\Delta\gamma_{\infty,i,j,k}$ (in other words
, each source, though close to each other, do not exactly cross the
potential at the same 
 position),
 \item the photometric redshifts are not the true redshifts and lead to shifts
$\Delta\omega
_{i,j, k}$ on the lensing factors.
\end{enumerate}
From equations  (13) it is clear that, due to the 
linearity of the 3-3 determinant and of the averaging on all the triplets,
the final noise 
is linear regarding to each individual term : it is composed to first order
of linear 
combination of terms (like $\bar{\epsilon}_{S,i}$),  and to second order of
linear 
combination of crossed-terms (like $\epsilon_{S,i}.\Delta\omega_{j}$).
 These crossed-terms  cannot introduce any systematic bias on G and thus 
can be neglected.\\
The following equations give the four different noise contributions: 
$GN_{\epsilon_{S,i}}$, due to the corrected source ellipticity, 
$GN_{\Delta\epsilon_{i}}$, due to the error on the ellipticity, 
$GN_{\Delta\omega_{i}}$ due to the approximate redshift and 
$GN_{\Delta\gamma_{\infty,i}}$ due to the potential,
%%%%%%%%%%%%%%%%%%%%%%%%%%%%%%%%%%%%%%%%%%
\begin{eqnarray}
  GN_{\bar{\epsilon}_{S,i}} \simeq <\gamma_{\infty}^{2}>
&< \omega_{j} \omega_{k}^{o} (\omega_{i}-\omega_{k}) \bar{\mbox{\boldmath 
$\epsilon$}}_{S,i}^{*} /\mbox{\boldmath $\gamma$}^{*} \\
\nonumber &+ \omega_{j}^{o} \omega_{k} (\omega_{j}-\omega_{i}) 
\bar{\mbox{\boldmath $\epsilon$}}_{S,i} /\mbox{\boldmath $\gamma$} > \\
  GN_{\Delta\epsilon_{i}} \simeq <\gamma_{\infty}^{2}>
&< \omega_{j} \omega_{k}^{o} (\omega_{i}-\omega_{k}) \Delta\mbox{\boldmath 
$\epsilon$}_{i}^{*} /\mbox{\boldmath $\gamma$}^{*} \\
\nonumber &+ \omega_{j}^{o} \omega_{k} (\omega_{j}-\omega_{i}) 
\Delta\mbox{\boldmath $\epsilon$}_{i} /\mbox{\boldmath $\gamma$} > \\
  GN_{\Delta\omega_{i}} \simeq <\gamma_{\infty}^{2}>
&< \Delta\omega_{i} (\omega_{j}\omega_{k}^{o} (\omega_{k}-\omega_{j}) \\
\nonumber &+  \omega_{j}\omega_{k}^{o}\omega_{i}^{o} - 
\omega_{j}^{o}\omega_{k}\omega_{i}^{o}) > \\
  GN_{\Delta\gamma_{\infty,i}} \simeq <\gamma_{\infty}^{2}>
&< \omega_{j}\omega_{k}^{o}(\omega_{i}-\omega_{k}) \Delta\mbox{\boldmath 
$\gamma$}_{i}^{*}/\mbox{\boldmath $\gamma$}^{*} \\
\nonumber &+ \omega_{j}^{o}\omega_{k}(\omega_{j}-\omega_{i}) 
\Delta\mbox{\boldmath $\gamma$}_{i}/\mbox{\boldmath $\gamma$}  > .
\end{eqnarray}
%%%%%%%%%%%%%%%%%%%%%%%
The terms of index $j$ and $k$ can be easily derived from these equations
by  cyclic 
permutations of the ($ijk$) triplet.\\
It is worth noting that even if the bias, as $G^{main}$, is a 
function of the cosmological parameters, we can neglect this dependency.
Indeed, for 
a given shift of $(\Omega,\Lambda)$ around $(\Omega_{o},\Lambda_{o})$ the 
variations of $G^{main}$  and $GN$ verify $\Delta R_e(GN)/\Delta G^{main} 
\propto 1/\sqrt{N}$.
Besides, provided that $N$ is large enough, the probability distribution of
the noise 
$R_e(GN)$ can be considered as a Gaussian law centered on 0 (as the sum of
a large   
number of nearly Gaussian laws). So far, we have shown that the variance of
this noise 
 decreases as  $1/\sqrt{N}$ because of the redundancy of triplets. This
applies for 
$GN_{\bar{\epsilon}_{S,i}}$, $GN_{\Delta\epsilon_{i}}$ and 
$GN_{\Delta\omega_{i}}$ but not for $GN_{\Delta\gamma_{\infty,i}}$. Indeed,
for 
a  galaxy $i$ included in two different triplets the associated term 
$\Delta\mbox{\boldmath $\gamma$}_{i}$ is different in each triplet. That is
why 
the variance of the fourth noise $GN_{\Delta\gamma_{\infty,i}}$ vanishes in 
$1/\sqrt{3 N_{tr}}$, where $N_{tr}$ is the total number of triplets :
$N_{tr}\approx 3N$ (see 
the remark of section 3.1).\\
The variances of the four noises behave approximately as follows (under the 
conditions that will be detailed in section 6) :
%%%%%%%%%%%%%%%%%%%%%%%%%%%%%%%%%%%%%
\begin{eqnarray}
   GN_{\bar{\epsilon}_{S,i}} &\approx& 0.08
\frac{<\gamma_{\infty}^{2}>}{\sqrt{N}} \ ,
\\
   GN_{\Delta\epsilon_{i}} &\approx& 0.02
\frac{<\gamma_{\infty}^{2}>}{\sqrt{N}}  \ , 
\\
   GN_{\Delta\omega_{i}} &\approx& 0.04  
\frac{<\gamma_{\infty}^{2}>}{\sqrt{N}} \ , \\
   GN_{\Delta\gamma_{\infty,i}} &\approx& 0.02 
\frac{<\gamma_{\infty}^{2}>}{\sqrt{N}} \ ,
\end{eqnarray}
%%%%%%%%%%%%%%%%%%%%%%%%%%%%
%%%%%%%%%%%%%%%%%%%%%%%%%%%%%%%%%%%%%%%%%%%%%
\subsection{Resulting signal to noise ratio}
Let us establish the relation between the signal to noise ratio 
of the method and the number of background galaxies (or the number of
considered 
clusters). In the following we will take only $1000$ galaxies per cluster
(see section 6).  If we define the signal as
a variation of $G^{main}$ along its gradient and $\sigma_{GN}$ as the
variance of the 
 statistical noise (see section 3.4),  the signal to noise ratio can thus
be written as :
\begin{eqnarray}
  S/N=\frac{\Delta G^{main}}{\sigma_{GN}}=\frac{\Delta_{\Omega,\Lambda} \left |
\left | 
\vec{\bigtriangledown}G^{main}_{\mid\Omega_{o},\Lambda_{o}}\right | \right
|}{\sigma_{GN}}.
\end{eqnarray}
The $\left | \left |...\right | \right |$ notation indicates the norm of a
vector. $\Delta_{\Omega,\Lambda}$ is the  accuracy on the cosmological
parameters.
Figure (3) plots the variation of $\Delta_{\Omega,\Lambda}$ as a function of
 the number of clusters (with an observed number density of background
galaxies equal to $50$ by square-arcminute) for a $1\sigma$ confidence
level. This plot shows that an accuracy of $0.1$ can be 
reached  (at a $1\sigma$ level) with about 1000 clusters, or 
similarly an accuracy of about $0.3$ can be reached  (at a $1\sigma$ level)
with $100$ clusters. \\
It shows that with a good seeing one could in principle test the existence
of a 
cosmological constant from VLT observations with about 100 clusters. Far
better it should be possible to measure the curvature of the Universe with
a reasonable accuracy from NGST observations (the number density of
background 
galaxies is multiplied by 10). These prospects of course require that the
systematics are previously corrected.
%%%%%%%%%%%%%%%%%%%%%%%%%%%%%%
\begin{figure}
\begin{center}
\psfig{figure=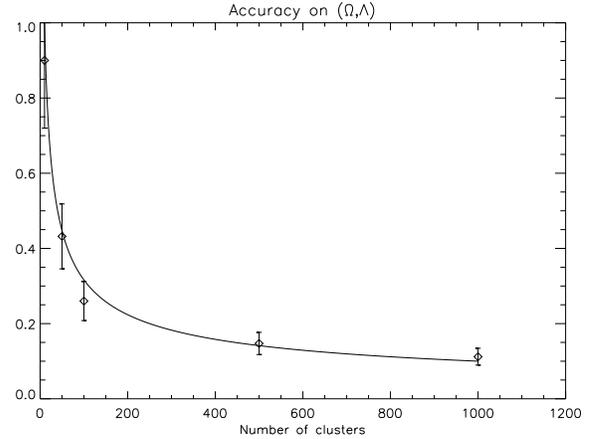,height=6cm}
\caption{Accuracy on the cosmological parameters  versus the number of
clusters (with $1000$ galaxies per cluster) at a $1\sigma$ confidence
level. The  plain line extrapolates the results obtained from simulations with an
inverse square root function.  Error bars account for the number of simulations.
We  can read that a $0.3$ accuracy (at
$1\sigma$) can be achieved with the observations of $100$ clusters or  a
$0.1$ accuracy (at $1\sigma$) can be achieved with the observations of
$1000$ clusters. Equivalently a $0.2$ accuracy (at $2\sigma$) can be
achieved with the observations of $1000$ clusters.}
\end{center}
\end{figure}
%%%%%%%%%%%%%%%%%%%%%%%%%%%%%
These results can be refined by the introduction of a probability
distribution.
%%%%%%%%%%%%%%%%%%%%%%%%%%%%%%%%%%%%%%%%%%%%%%%%%%%%%%%%%%%%%%%%
\subsection{The $(\Omega,\Lambda)$ probability distribution}
Thanks to the Gaussian behavior of the noise we can derive from the operator 
$G(\Omega,\Lambda)$ a probability distribution $p(\Omega,\Lambda)$ for the
cosmological parameters :

\begin{equation}
\displaystyle{    p_{\Omega,\Lambda}(\Omega,\Lambda)} \equiv
\displaystyle{
 {\rm exp} 
  \left (- 
   {
   \left[ R_e
    \left[ G
     \left (\Omega,\Lambda
     \right )
    \right ] 
   \right ]^2 
\over 2 \sigma_{GN}^{2}
   }
  \right )
}
 \  ,
\end{equation}
%%%%%%%%%%%%%%%%%%%%%%%%
which has just to be normalized on the considered domain of 
$(\Omega,\Lambda)$, (for example $\Omega+\Lambda=1$).\\
The operator $G(\Omega,\Lambda)$ is directly derived from the 
observational data. On the other hand, 
 $\sigma_{GN}$ is obtained from simulations applied on the data in 
the following way. Once  the 
function $G(\Omega,\Lambda)$ is obtained from the observation of
($\mbox{\boldmath 
$\epsilon$}$, z), it is possible to add noise to   
  the data by adding random 
 ellipticities ($\mbox{\boldmath $\epsilon$}_{S}$) and  
 $\Delta z$ random errors to the redshifts $z$ : the variance of the noise
$GN$ 
induced on $G$ is then directly $\sigma_{GN}$.
%%%%%%%%%%%%%%%%%%%%%%%%%%%%%%%
\section{Analysis of systematics}
In this section we discuss the systematics which could potentially reduce 
significantly  the efficiency of the method from an observational point of
view.  Though we give estimations of their amplitude,  it is worth noting
that 
a reliable quantitative estimate of their impact on the triplet method will
demand additional work (mainly simulations and calculations taking in
account the perturbation coming from large-scale structures. 
This effect will be studied in paper II).    \\
Two main systematics have been identified:
\begin{enumerate}
\item a  systematic bias on $G$ produced by to the non-symmetry of the term
$\Delta\omega$. 
It can be easily corrected (see 4.1). \\
\item a contamination produced by  the existence of background structures
which play the role 
of perturbing lenses. They can be  the background galaxies themselves
(which induce galaxy-galaxy lensing) and possibly any condensation of mass
(clusters or large-scale structures) located on the  line of sight of the
lensing cluster. In the following sections  (section 4.2 and 4.3), we focus
on  two extreme 
cases: the  systematic  produced by the  galaxies of each triplet
themselves, and 
 those generated by large-scale structures. This study will be performed
using multi-lensing 
models (see Kovner 1987). Appendix B gives the calculation of the measured
ellipticities in a multi-lensing model.
\end{enumerate}
%%%%%%%%%%%%%%%%%%%%%%%%%%%%%%%%%%%%%%%%%%%%%
\subsection{Non symmetry of the term $\Delta\omega$}
The method supposes that we know the photometric redshifts of background
galaxies 
 as well as the probability distribution $p_{\Delta z}(\Delta z)$ of the
error between the 
photometric redshift and the true one. Even if $p_{\Delta z}(\Delta z)$ 
is symmetric and centered on 0, which is not necessarily the case, the
probability 
distribution of the resulting  shift on $\omega$, 
$p_{\Delta\omega}(\Delta\omega)$ may be non symmetric and the mean value of
$\Delta\omega$ 
may be different from zero, introducing a systematic in equation (23).
Fortunately, this 
effect is easy to correct. Since the redshift distribution 
$n(z)$ is known the systematic can be exactly balanced by replacing in the
definition 
of $G_{ijk}$ $\omega$ by $\omega-\overline{\Delta\omega}$, where
\begin{eqnarray}
   \overline{\Delta\omega}(z)=\int \omega(z+\Delta z)n(z+\Delta z)p_{\Delta 
z}(\Delta z) d\Delta z   \ .
\end{eqnarray}
%%%%%%%%%%%%%%%%%%%%%%%%%%%%%%%%%%%%%%%%%%%%%%%%%%%%%%%%%%
\subsection{Potential perturbation due to galaxy-galaxy lensing}
Since we consider triplets of galaxies having different redshifts
 but which are very close together in the image plane,  the most distant  
  can potentially be distorted by the lensing effect of the others.
 In particular, in each triplet 
($i, j, k$) (ordered with increasing redshifts) the galaxy $i$ is a
perturbation of the potential 
for the sources $j$ and $k$ as well as the galaxy $j$ for the source $k$.
It is 
worth noting that $i$
 is not the only perturbing lens for the source $j$, and  
$i$ and $j$ are not the only one for 
  the source $k$. For instance, each galaxy may be itself part of a 
group of galaxies which  produces its own  additional lensing effect.
 Nevertheless, in order to get a rough estimate of the amplitude of this
kind of
systematics,  we will assume to first approximation
 that the galaxy-galaxy lensing produced within each triplet,  
is the dominant contribution. \\ 
The galaxy-galaxy lensing  increases as the angular distance between the 
galaxy-lens and the galaxy-source decreases. 
 It also depends on their relative redshifts: that is why it induces a
systematic bias 
on the construction of $G$. In Appendix B we compute the amplitude of the
corrections on the 
 measured ellipticities once 
 several lenses along the line of sight are taken into account.
 From Equations (62), (63) and (66) of Appendix B, one can see that these 
 perturbations produce two kinds of taxes:
\begin{enumerate}
\item A purely additive term, $g^{P}$. It can be seen as the linear
contribution from the shear, as if the shears of each lenses were simply
added to each other without coupling considerations.
\item A a multiplicative term of the form
$\left(1-\left(1-c\right)\kappa^{P}\right)^{-1}$ 
 which modifies the main cosmological term $g$. It accounts for couplings
between the main and additional lenses.
\end{enumerate}
%%%%%%%%%%%%%%%%%%%%%%%%%%%%%%%%%%%%%%%%%%%%%%%%%%
\subsubsection{The additive linear term.}

Since the three galaxies of each triplet are randomly distributed and
uncorrelated, this term induces a noise rather than a systematic bias. It
decreases as $1/\sqrt{3N}$, like the noise $GN_{\Delta\gamma_{\infty}}$ 
for instance. 
Besides, the shear induced by a galaxy-lensed can be controlled (if we
cancel the triplets where the angular distance between two of the three
sources is too small: typically lower then a few arcsecond for typical
Einstein radius of galaxies. This means an induced shear inferior to about
$2\%$). Therefore the amplitude of this additive term can  be bounded to
less than $10\%$ of the effect coming from the principal statistical noise
$GN_{\epsilon_{S,i}}$, and 
thus can be neglected. \\
%%%%%%%%%%%%%%%%%%%%%%%%%%%%%%
\subsubsection{The multiplicative coupling term.}
The effect of coupling between two lenses depends on their relative
redshifts through angular distances ratio, hence this term really induces a
systematic bias on $G$. To calculate it we have to think about the meaning
of the additional convergence, $\kappa^{P}$, used in the multi-lens screen
approach of the annex B.  This convergence associated to the source $j$
and   
 produced by the galaxy-lens $i$ will be noted $\kappa^{i,j}$; the
convergences 
associated to the source $k$ and produced by the galaxy-lenses $i$ and $j$ 
 will be noted  $\kappa^{i,k}$ and $\kappa^{j,k}$. We cannot directly
compare their value since 
they depend on the redshifts $z_{i},z_{j},z_{k}$. So, in order to model this 
systematic bias we associate to each of these convergences an absolute one 
$\tilde{\kappa}^{gal}$ independent of the redshifts $z_{i}$, $z_{j}$ and
$z_{k}$ 
(obviously $\tilde{\kappa}^{gal}$ should depend on the angular distance
between the
galaxy-source and the galaxy lens. We consider here a mean value averaged
over the area around 
each galaxy-lens where we search for a source. For the definition of this
area, see section 5.2):
%%%%%%%%%%%%%%%%%
\begin{eqnarray}
 \kappa^{i,j} &=&   
\frac{D_{Oi}}{D_{OL}}\frac{D_{ij}}{D_{Oj}}\tilde{\kappa}^{
gal}\\ 
 \kappa^{i,k} &=& 
\frac{D_{Oi}}{D_{OL}}\frac{D_{ik}}{D_{Ok}}\tilde{\kappa}
^{gal}\\
 \kappa^{j,k} &=& 
\frac{D_{Oj}}{D_{OL}}\frac{D_{jk}}{D_{Ok}}\tilde{\kappa}
^{gal}. 
\end{eqnarray}
%%%%%%%%%%%%%%%%%%
The $\Delta_{\Omega,\Lambda}^{GGL}$ shift (along the gradient of $G$) on
the cosmological parameters determination and due to the galaxy-galaxy
lensing perturbation effects is then  :
%%%%%%%%%%%%%%%%%%%%%%%%
\begin{eqnarray}
  \frac{\left <\left| \begin{array}{lcr}
      1 & \omega_{i} & 0 \\
      1 & \omega_{j} &
\frac{D_{ij}}{D_{OL}}(\frac{D_{Oi}}{D_{Oj}}-\frac{D_{Li}}{D_{Lj}}) 
\\ 
      1 & \omega_{k} & \frac{D_{ik}}{D_{OL}}(\frac{D_{Oi}}{D_{Ok}}-
\frac{D_{Li}}{D_{Lk}})+\frac{D_{jk}}{D_{OL}}(\frac{D_{Oj}}{D_{Ok}}-
\frac{D_{Lj}}{D_{Lk}})\\
                 \end{array}  \right| \right >} {\left | \left | 
\vec{\bigtriangledown}_{\Omega_{0},\Lambda_{0}}\left <\left|
\begin{array}{lcr}
      1 & \omega_{i} & \omega_{i}/\omega_{i}^{o} \\
      1 & \omega_{j} & \omega_{j}/\omega_{j}^{o} \\ 
      1 & \omega_{k} & \omega_{k}/\omega_{k}^{o}\\
                 \end{array}  \right| \right > \right | \right |}
\tilde{\kappa}^{gal} 
\end{eqnarray}
%%%%%%%%%%%%%%%
\noindent where the brackets  $\left <...\right >$ denote  the average over
all the ordered triplets. \\
Using the redshift distribution (42) and considering the mean value of the
convergence
created by a galaxy as about $2\%$ (see section 5.2), one gets :
$\Delta_{\Omega,\Lambda}^{GGL}=10.2\tilde{\kappa}^{gal} \approx 0.2$. \\
This is only an indicative value. However, with galaxy/galaxy lensing
measurements,
we are in principle able to account for the effect of the potential
of each background galaxy. This systematic can be
calculated and corrected. Therefore, if the knowledge of the mean potential
of galaxies can be reached with about a $\approx20\%$ accuracy (we could
use Faber-Jackson models or wait for the incoming results of  galaxy-galaxy
lensing studies (see Schneider et al. 1996)), the remaining systematic
would then be about $\Delta_{\Omega,\Lambda}^{GGL}\approx 0.04/\sqrt{Nclust}$.
\\
This correction does not take in account the possibility that a fraction of
the 
 galaxies are embedded in  groups which could enhance the contamination. 
There are two possibilities to solve this issue. Since we know the
photometric redshifts of 
the galaxies we can discriminate compact groups of galaxies and remove the
corresponding 
triplets from the sample. Or, if the galaxy-galaxy lensing results are able
to account for these 
effects with a reasonable accuracy, then we can as previously calculate
$\Delta_{\Omega,\Lambda}^{GGL}$. \\
In conclusion this effect is non-negligible but can be accurately
estimated and corrected, thanks 
to the forthcoming observations of galaxy-galaxy lensing on blank fields.
%Alternatively, one could even imagine  
%that the galaxy-galaxy lensing effects used jointly with the knowledge of color
%redshifts,  
%and applied to the huge number of galaxies that will be found on wide field
%survey,  can possibly be an another way to constrain 
%the cosmological parameters. 
%%%%%%%%%%%%%%%%%%%%%%%%%%%%%%%%%%%%%%%%%%%%%%%%%%%%%%%
\subsection{Potential perturbation due to background lensing structures}
Let us consider the perturbation  due to  background structures. Before
giving an estimate of it
for the case of matter structures integrated along the line of sight, we
calculate the perturbation
due to a single lens plane (containing an over dense region like a 
cluster or even an under dense region). Since this second lens may be
located anywhere in redshift  then it differently affects the measured
ellipticities of the sources ($i,j,k$) 
in each triplet, it can induce a systematic on the value of $G$. Once again
and still using 
the calculations done in annex B (see equation (66)), this systematic can
be split in 
two parts due to an additive term (corresponding to the linear contribution
of the perturbative lens) and a multiplicative term (corresponding to the
coupling between the main and the perturbative lenses).
%%%%%%%%%%%%%%%%%%%%%%%%%%%%%%%%%%%%%%%%%%%%%%%%%%%%
\subsubsection{The additive linear term.}
Like in the above section, we associate an absolute convergence and shear 
$\delta\tilde{\kappa}$ and $\delta\tilde{\gamma}$ to the 
additional convergence and shear $\delta\kappa^{P,S}$ and
$\delta\gamma^{P,S}$ coming 
from the lensing effect of the structure on a source S (S can be $i$, $j$
or $k$):
%%%%%%%%%%%%%%%%%%%%%%%%%%%%%%%%%%%%%%%%%%
\begin{eqnarray}
\delta\kappa^{P,S} &=& 
\frac{D_{OL^{P}}}{D_{OL}}\frac{D_{L^{P}S}}{D_{0S}}\ \delta\tilde{
\kappa} \\
\delta\gamma^{P,S} &=& 
\frac{D_{OL^{P}}}{D_{OL}}\frac{D_{L^{P}S}}{D_{OS}}\ \delta\tilde{
\gamma} \ ,
\end{eqnarray}
%%%%%%%%%%%%%%%%%%%%%%%%%%%%%%%%%%%
\noindent where the superscript $^P$ denotes the perturbative term. \\
We assume that the convergence and the shear are constant all over 
the image. The $\Delta_{\Omega,\Lambda}^{LBS}$ shift (along the gradient of
$G$) on the cosmological parameters determination and due to the linear
lensing effect of the background structure is then:
%%%%%%%%%%%%%%%%%%%%%%%%%%%%%%%%%%%%%%%%%
\begin{eqnarray}
 \Delta_{\Omega,\Lambda}^{LBS} =\frac{\left <\left| \begin{array}{lcr}
      1 & \omega_{i} & \frac{D_{L^{P}j}}{D_{Lj}}\frac{D_{L^{P}k}}{D_{Lk}}  \\
      1 & \omega_{j} & \frac{D_{L^{P}k}}{D_{Lk}}\frac{D_{L^{P}i}}{D_{Li}} \\ 
      1 & \omega_{k} & \frac{D_{L^{P}i}}{D_{Li}}\frac{D_{L^{P}j}}{D_{Lj}} \\
                 \end{array}  \right| \right >\left (
\displaystyle{{D_{OL^{P}}\over {D_{OL}}}
}
\right)^{2} } {
\left | \left | \vec{\bigtriangledown}_{\Omega_{0},\Lambda_{0}}\left <\left| 
\begin{array}{lcr}
      1 & \omega_{i} & \omega_{i}/\omega_{i}^{o} \\
      1 & \omega_{j} & \omega_{j}/\omega_{j}^{o} \\ 
      1 & \omega_{k} & \omega_{k}/\omega_{k}^{o}\\
                 \end{array}  \right| \right > \right | \right |} 
\left (\frac{\delta\tilde{\gamma}}{\gamma_{\infty}}\right )^{2}.
\end{eqnarray}

For a condensation of mass located at redshift $1$ and with the redshift
distribution (43) this systematic 
is : $\Delta_{\Omega,\Lambda}^{LBS}=1.8\left
(\frac{\delta\tilde{\gamma}}{\gamma_{\infty}}\right )^{2}$.  To generalize
this systematics to large scale structures, the above value of
$\Delta_{\Omega,\Lambda}^{LBS}$ has to be integrated along the line of
sight. Here we do not perform exactly this calculation
(which needs the introduction of cosmological scenario and the non linear
evolution of perturbations) but only give an estimation of it using the
already known results from the  studies of lensing by large scale
structures (see the review by Mellier 1998). 
Calculations from the non linear evolutions of the power spectrum in the
one minute angular scale predict a  polarization of about $3\%$.
From this value  we can make a rough estimate of the shift
on $(\Omega,\Lambda)$ (using the same conditions as will be used in the
simulations, section 6) : $\Delta_{\Omega,\Lambda}^{LBS}\approx 0.03$.\\
This is only an indicative value, however this systematic can be estimated
more accurately using the simulations of the non-linear evolution of
perturbations.
% The  aim of the next coming paper will be to test directly
%from simulations the perturbation effect coming from large scale structures
%and compare its amount with what can be calculated with the non linear
%evolution of perturbations.
%%%%%%%%%%%%%%%%%%%%%%%%%%%%%%%%%%%%%%%%%%%%%%%%%%%
\subsubsection{The multiplicative coupling term.}
The $\Delta_{\Omega,\Lambda}^{CBS}$ shift (along the gradient of $G$) on
the cosmological parameters determination and due to the coupling  effect
of a background perturbing structure (with the main lens) is calculated  as
in 4.2.2 but  
is simpler since there is only one perturbating lens to which we associate
an absolute 
convergence and shear (see equation (66)). It leads to:
%%%%%%%%%%%%%%%%%%%%%%%%%%%%
\begin{eqnarray}
 \Delta_{\Omega,\Lambda}^{CBS} =\frac{\left <\left| \begin{array}{lcr}
      1 & \omega_{i} & \frac{D_{L^{P}i}}{D_{OL}}(\frac{D_{OL^{P}}}{D_{Oi}}-
\frac{D_{LL^{P}}}{D_{Li}}) \\
      1 & \omega_{j} & \frac{D_{L^{P}j}}{D_{OL}}(\frac{D_{OL^{P}}}{D_{Oj}}-
\frac{D_{LL^{P}}}{D_{Lj}})\\ 
      1 & \omega_{k} & \frac{D_{L^{P}k}}{D_{OL}}(\frac{D_{OL^{P}}}{D_{Ok}}-
\frac{D_{LL^{P}}}{D_{Lk}})\\
                 \end{array}  \right| \right >} {\left | \left | 
\vec{\bigtriangledown}_{\Omega_{0},\Lambda_{0}}\left <\left|
\begin{array}{lcr}
      1 & \omega_{i} & \omega_{i}/\omega_{i}^{o} \\
      1 & \omega_{j} & \omega_{j}/\omega_{j}^{o} \\ 
      1 & \omega_{k} & \omega_{k}/\omega_{k}^{o}\\
                 \end{array}  \right| \right > \right | \right |}
\  \delta\tilde{\kappa}.
\end{eqnarray}
The same discussion as above applies. For a condensation of mass located at
redshift $1$ and with the redshift distribution (43) this systematic is : 
$\Delta_{\Omega,\Lambda}^{CBS}=16.7\delta\tilde{\kappa}$.  To generalize
this systematic to large scale structures, the above value of
$\Delta_{\Omega,\Lambda}^{CBS}$ has to be integrated along the line of
sight. We use again a polarization of about $3\%$. Here the difference is
that this systematic effect
behaves like a statistic noise because the $\delta\tilde{\kappa}$ can be
either positive or negative. Hence,
when the triplet method is applied on many different clusters, we expect
the averaged value of this 
systematic to behave roughly as the inverse of the square root of the
number of considered clusters.
As an example, for $100$ clusters, we expect this systematic to be about
$\Delta_{\Omega,\Lambda}^{CBS}\approx 0.04$. \\
Contrary to the other two systematics ($\Delta_{\Omega,\Lambda}^{GGL}$ and
$\Delta_{\Omega,\Lambda}^{LBS}$), this one can not be corrected because of
its random behavior. However its effect on the $(\Omega,\Lambda)$
determination decreases with the number of considered clusters.

\subsubsection{Effect of a background cluster}
In the last two sections we discussed the non-linear evolution of
background structures in a 
scale of about
one arcminute to estimate the resulting systematics on the triplet method. 
It may be also important to  take in account (statistically) the effect of
background clusters.
 In particular,  the accidental presence of another  cluster  
 lying along  the line of sight of the main lensing-cluster could be a
serious artefact.
 In the case of usual clusters of galaxies, these biased targets  can be
easily removed from the
sample by using photometric redshift informations.
 However, if it happens that dark clusters do really exist,
then their presence could be extremely difficult to detect. 
Their impact on the statistics
depend on the number density of such systems, if any.  Since we do not have
any evidence 
 from shear map that dark clusters exist, this is a somewhat an academic
problem. However, such structures could be revealed 
 by using a
method like the aperture mass density. In principle, this technique is 
 able to detect a dark halo, provided that their velocity dispersion is
larger than $400$ km/s (see Schneider et al. 1995).\\
A detailed quantitative estimate of these effects require simulations and
also additional
 observations in order to constrain the fraction of lensing clusters which 
 is contaminated by another cluster along the line of sight.  This
investigation 
 is beyond the scope of this paper, but should be addressed in a
forthcoming paper.
%%%%%%%%%%%%%%%%%%%%%%%%%%%%%%%%%%%%%%%%%
\subsection{Outcome}
Estimations of the various systematic biases are given in the following
table, for a principal lens at redshift $0.4$, and a redshift distribution
similar to (42): \\
%%%%%%%%%%%%%%%%%%%%%%%%%%%%%%%
\begin{table}
\begin{center}
\begin{tabular}{|c|c|c|}
\hline
 & & \\
$\Delta_{\Omega,\Lambda}^{GGL}$ & $\approx 0.2$ & corrected to $\approx
0.04/\sqrt{N_{clust}}$\\
 & & \\
\hline 
 & & \\
$\Delta_{\Omega,\Lambda}^{LBS}$ & $\approx 0.03$ & corrected \\
 & & \\
\hline 
 & & \\
$\Delta_{\Omega,\Lambda}^{CBS}$ & $\approx 0.4/\sqrt{N_{clust}}$ & not
corrected\\
 & & \\
\hline 
\end{tabular}
\caption{The systematic biases on the determination of $(\Omega,\Lambda)$.
$\Delta_{\Omega,\Lambda}^{GGL}$ is due to galaxy-galaxy lensing effects.
$\Delta_{\Omega,\Lambda}^{LBS}$ is due to the linear effect of the
distortion induced by background
structures. $\Delta_{\Omega,\Lambda}^{CBS}$ comes from the coupling
of these background structures with the main lens.The corrections of 
$\Delta_{\Omega,\Lambda}^{GGL}$ and $\Delta_{\Omega,\Lambda}^{LBS}$ require
to adjust the method with a modeling  of the potential of galaxies and
large scale structures. Such modeling compared with ray tracing simulations
will be performed in paper II. $\Delta_{\Omega,\Lambda}^{CBS}$ can not be corrected
because of its random behavior.}
\end{center}
\end{table} 
Next section will give qualitative solutions to deal with part of these
systematics and to increase the signal to noise ratio of the method.\\ 
%%%%%%%%%%%%%%%%%%%%%%%%%%%%%%%%%%%%%%%%%%%%%%%%%%%%%%%%%%%%%%%%%%%%%%%%%%%%%%
\section{Optimization of the method.}
 So far, the method has been presented in a structure as general as
possible, with no 
particular choice for the geometry of clusters, neither for the redshifts
nor for 
 the distances between galaxies inside triplets. This section discusses
qualitatively 
 these degrees of freedom in order to increase the signal and decrease 
the bias $\Delta_{\Omega,\Lambda}^{LBS}$ in the same time.
%%%%%%%%%%%%%%%%%%%%%%%%%%%%%%%%%%%%%%%%%%%%%%%%%%%%%%%%%%%%%%%

\subsection{Choosing clusters with symmetrical geometry}
The method as explained so far can be applied to every cluster 
 regardless its  geometry. 
However new cluster samples, like the one  obtained by the X-ray satellite
XMM,  
 will be 
 useful to select only those with a symmetrical geometry, since any
geometrical information 
can be used to decrease the systematic bias noticed due to the linear
perturbation effect of background structures  (see section 4.3.1). This bias 
is due to scalar products as $\mbox{\boldmath $g$}^{P,i}\mbox{\boldmath
$g$}^{P,j*}$ 
appearing from the construction of $G_{ijk}$ in the determinant:
%%%%%%%%%%%%%%%%%%%%%%%%%%%%%%%%%%%%%%%%%%%%%%%%%%%%
\begin{eqnarray}
   \left| \begin{array}{lcr}
      1 & \omega_{i}^{o} & \omega_{i}^{o}\mbox{\boldmath
$g$}^{P,j}\mbox{\boldmath $g$}^{P,k*} \\
      1 & \omega_{j}^{o} & \omega_{j}^{o}\mbox{\boldmath
$g$}^{P,k}\mbox{\boldmath $g$}^{P,i*} \\ 
      1 & \omega_{k}^{o} & \omega_{k}^{o}\mbox{\boldmath
$g$}^{P,i}\mbox{\boldmath $g$}^{P,j*} \\
              \end{array}  \right| . 
\end{eqnarray}
%%%%%%%%%%%%%%%%%%%%%%%%%%%%%%%%%
Assume that we apply the method to a circular cluster. Then we can construct 
 a sub-sample of triplets of galaxies  inside a ring  
centered on the cluster center. Each galaxy ($i$, $j$ or $k$) of the
triplet is 
associated with an angle, $\alpha_{i},\alpha_{j},\alpha_{k}$. Hence, we can
replace each 
measured ellipticity ($\mbox{\boldmath $\epsilon$}_{i}$,  
$\mbox{\boldmath $\epsilon$}_{j}$ and $\mbox{\boldmath $\epsilon$}_{k}$) by
the 
tangential ellipticities ($\mbox{\boldmath $\epsilon$}_{i}e^{2i\alpha_{i}}$,  
$\mbox{\boldmath $\epsilon$}_{j}e^{2i\alpha_{j}}$ 
and $\mbox{\boldmath $\epsilon$}_{k}e^{2i\alpha_{k}}$), the above
determinant is replaced by:
%%%%%%%%%%%%%%%%%%%%%%%%%%%
\begin{eqnarray}
   \left| \begin{array}{lcr}
      1 & \omega_{i}^{o} & \omega_{i}^{o}\mbox{\boldmath
$g$}^{P,j}\mbox{\boldmath $g$}^{P,k*}e^{2i(\alpha_{j}-\alpha_{k})} \\
      1 & \omega_{j}^{o} & \omega_{j}^{o}\mbox{\boldmath
$g$}^{P,k}\mbox{\boldmath $g$}^{P,i*}e^{2i(\alpha_{k}-\alpha_{i})} \\ 
      1 & \omega_{k}^{o} & \omega_{k}^{o}\mbox{\boldmath
$g$}^{P,i}\mbox{\boldmath $g$}^{P,j*}e^{2i(\alpha_{i}-\alpha_{j})} \\
              \end{array}  \right| .
\end{eqnarray}
%%%%%%%%%%%%%%%%%%%%%%%%%%%%%%%%%%%%%%%%%
We can see that the arguments of the elements of the third column are
randomly 
distributed. Therefore the systematic bias $\Delta_{\Omega,\Lambda}^{LBS}$
vanishes as the inverse of the square root of the number of triplets and
becomes negligible, whereas the main term is not modified.\\
Actually the signal to noise ratio of the method (as defined in (29)) can
also be increased 
in the case of a circular geometry because the number of triplets becomes
large enough to 
enable a stringent selection within them: either by the angular environment
of the 
three sources (see 5.2), either by their redshift (see 5.3.1). 
%%%%%%%%%%%%%%%%%%%%%%%%%%%%%%%%%%%%%%%%%%%%%
\subsection{Choosing the triplets of galaxies}
This section refers to the necessity of rejecting of the sample
  sources for which another galaxy may play 
the role of a galaxy lens. The angular radius of the circle in which we can 
consider that a galaxy is isolated without important local perturbation of
another 
galaxy depends on the real mass distribution of galaxy halos.  Therefore 
 it is somewhat difficult to provide accurate estimate of this circle with
our present-day
knowledge of the distant galaxy halos. As a rough 
simplification we can consider each galaxy as an isothermal sphere and then
choose the
 angular radius such that the  induced shear (and convergence) is lower
than   
 $5\%$. This conservative approximation applies not only in the section 
4.2.1, but also in 4.2.2, where $\tilde{\kappa}^{gal}$ can thus be chosen
small ($\approx 2\%$) in 
order to decrease the systematic bias $\Delta_{\Omega,\Lambda}^{GGL}$.\\
Of equal importance is the question of the maximum distance $\Delta r$
tolerable between 
each component of the triplet. Or, in the case of a circular cluster
analysis, what is the thickness
  of the rings ? To answer we need to balance two opposite effects: first, 
while $\Delta r$ increases, $<\gamma_{\infty}>$ decreases proportionally to
about 
$1-\Delta r/r$ (if the shear of the cluster is a law in $1/r$, as for an
isothermal sphere), 
so the signal to noise ratio decreases; second, in the same time the number
of triplets 
increases making possible a judicious selection of the redshifts in the
triplets 
(see section 5.3.1). After realistic simulations (using a number density of
galaxies 
equal to $50$ galaxies arcmin$^{-2}$), the optimized value that has been
selected 
for the first presentation of the method is $20$ arcseconds separation. In
practice, this can be optimized accordingly to the shear map of the cluster.
%%%%%%%%%%%%%%%%%%%%%%%%%%%%%%%%%%%%%%%%%%%%%%%%%%
\subsection{Redshift optimization}
In this section we suppose that the redshift distribution $n(z)$ and the 
variance of the redshift error $\sigma_{\Delta z}(z)$ are known. We then
investigate if 
the signal to noise ratio of the method can be improved thanks to either a
choice in the 
redshifts of the triplets or a selection in redshift of the clusters. In
all the oncoming considerations
the number density of background galaxies has been taken equal to $50$
galaxies arcmin$^{-2}$).
%%%%%%%%%%%%%%%%%%%%%%%%%%%%%%%%%%%%
\subsubsection{Redshifts in triplets}
From the matrix form (15), it appears obvious that if two of the three
redshifts of a triplet are very close together then the resulting value of
$G_{ijk}^{main}$ is nearly 
 zero and thus do no bring any cosmological information to $G$. 
Therefore it is necessary to set two minimum redshift differences 
$\Delta_{ij}=z_{j}-z_{i}$ and $\Delta_{jk}=z_{k}-z_{j}$ below which the
triplets 
will not be rejected. On the other hand, if 
these minima are too high then too many triplets will be rejected,
increasing 
the terms $1/\sqrt{N}$ and $\sigma_{GN}$. The right balance is obtained for 
$\Delta_{ij}$ and $\Delta_{jk}$ greater than about $0.05$. \\
This small value has been obtained from simulations. It  proves that in the
balance, 
the second effect is stronger. It means that with 
this density of galaxies, any stringent selection makes $\sigma_{GN}$
decrease rapidly. This 
will not be true anymore if we can select triplets within rings, for a
nearly circular 
cluster, or if we use observations with NGST (which will increase
significantly the number 
density of galaxies). In these two cases , the balance becomes favorable to
selections 
and so to an increase of the signal to noise ratio of the method.\\
The second concern about the redshift selection is  the difference between
$z_{i}$ and 
the redshift of the lens, $\Delta_{li}$. Once again, the optimization
arises with the balance 
of two competing effects : if $z_{i}$ is too close to $z_{l}$ then the
noise is increased 
by a nearly infinite value of $\Delta\omega_{i}$; however, in the same time
the 
gradient of $G_{ijk}^{main}$ is greater. Our simulations show that
  the right balance is obtained for $\Delta_{li}$ greater than about
$0.05$, and the same discussion as above applies. 
%%%%%%%%%%%%%%%%%%%%%%%%%%%%%%%%%%%%%%%%%%%%%%%
\subsubsection{Redshifts of clusters}
We can also find  the best balance between the last two effects when 
the redshift of the cluster changes. The most favorable cluster redshifts 
are strongly dependent on the shape of $\sigma_{\Delta z}(z)$ : the
redshift of the 
cluster must be in the area where the errors on the redshifts are as small
as possible 
for $\Delta\omega_{i}$ to be non dominant in the signal to noise ratio.\\
Besides, if the redshift of the cluster is too low then 
$\omega_{i}\approx\omega_{j}\approx\omega_{k}$ whatever the triplet is, the     
$S/N$ decreases.\\
With $\sigma_{\Delta z}(z)$ approximately equal to $0.03$ for redshift
inferior to 0.8 
and equal to $0.1$ for other redshifts, simulations show that clusters at
redshift between $0.3$ 
and $0.5$ are the most favorable to the method.
%%%%%%%%%%%%%%%%%%%%%%%%%%%%%%%%%%%%%%%%%%%%%
\section{Simulations and results}
In the simulations, we assumed that the redshift of each galaxy was known 
and we took a redshift distribution $n(z)$ which represents both the 
usual peak of galaxies near $z=1$ (described by the redshift distribution
$p(z)$ from 
Brainerd et al. 1996) 
 as well as a more distant population of faint blue galaxies suggested by
deep spectroscopic survey and that seem detectable in the selected color
bin $B=26-28$ (see Fort et al 1996, Broadhurst 98):
%%%%%%%%%%%%%%%%%%%%%%%%%%%%%%%%%%%%
\begin{eqnarray}
 n(z)=\alpha_{1}p(z)+\alpha_{2}p(z-2) \ \mbox{with} 
\end{eqnarray}
\begin{equation}
p(z)={
      \beta z^2 \over \Gamma\left(
      {3 \over \beta}
                            \right) z_0^3 } 
 {\rm exp}\left[
   -\left({z \over z_0}\right)^{\beta} \right]
    \ ,
\end{equation}
\noindent with $z_{0}=1/3$ and $\beta=1$, the $p(z)$ and $p(z-2)$
distributions give 
respectively an average redshift equal to 1 and 3.  
With $\alpha_{1}=0.7$ and $\alpha_{2}=0.3$, the fraction of high and low
redshift galaxies 
found by Fort et al  is respected. It  concerns a single observation of
very faint galaxies in B and I color through the cluster Cl0024+1654.
Although Broadhurst seems to have confirmed the result of this observation
with the Keck telescope, the redshift distribution may not be exactly
representative of the general redshift distribution of the faint galaxy
population for which we  can determine good color redshift. However, it is
important to remember that  the triplet method actually test the convexity
of  the lensing factor curve $\omega(z)$  (see Fig.1). Therefore, the
method mostly depends on the total number of galaxies above $z=1$ that we
can detect and for which we can determine an accurate color redshift (high
signal to noise for B photometry is essential). If a deep multicolor
photometry from U to K gives a broader redshift distribution with more
galaxies at intermediate redshift ($z=1.5-2$) the result of the triplet
method will be better since the new distribution will increase the number
of possible triplets with significant information on the cosmology. \\
We have also put realistic uncertainties in the photometric redshift of
galaxies. We assume it is  a Gaussian distribution with a redshift
dispersion equal to 0.03 for redshifts lower than 0.8 and as large as 0.1
for redshifts greater than 0.8 (see Brunner et al. 1997 and 
Hogg et al. 1998). \\
The intrinsic ellipticity distribution is chosen as follows :
%%%%%%%%%%%%%%%%%%%%%%%%%%%%%%%
\begin{eqnarray}
 p_{\mbox{\boldmath $\epsilon$}_{S}}(\mbox{\boldmath $\epsilon$}_{S}) = 
\frac{1}{\pi \sigma_{\epsilon_{S}}^{2} \left(1 -
e^{-1/\sigma_{\epsilon_{S}}^{2}}\right)} {\rm
exp}\left[-\left({\epsilon_{S}\over\sigma_{\epsilon_{S}}}\right)^{2}\right]
\end{eqnarray}
%%%%%%%%%%%%%%%%%%%%%%%%%%
\noindent with the intrinsic ellipticity dispersion $\sigma_{\epsilon_{S}}
= 0.15$ (from Seitz \& 
Schneider  1997).\\
For errors on the measured ellipticities, we chose the same distribution, with
  a dispersion equal to $0.02$.\\
For the clusters, the redshift have been taken equal to $0.4$, and their
projected mass
density  has the following analytical shape : 
%%%%%%%%%%%%%%%%%%%%%%%%%%%%%%%%%%%%%%%%%%
\begin{eqnarray}
   \kappa(r,\theta)={\kappa_{o} \over r\sqrt{1+e \cos{2\theta}}}.
\end{eqnarray}
%%%%%%%%%%%%%%%%%%%%%%%%%%%%%%%%%%%%%
The cluster ellipticities were chosen randomly between $0$ and $0.5$. It is
 unimportant to include a  core radius  since its influence in the weak
lensing area 
 is negligible  (Furthermore, we recall that the principle of the method is
to cancel 
any dependency on the mass map of the cluster). $\kappa_{o}$ is taken to
correspond to a sample of high  velocity dispersion ($1000$ km/s).\\
The size of the observation window has been taken similar to what will give
the VLT 
instrument FORS : $6'\times 6'$. Within this window, a circle of $90"$
radius representing 
the arclets and strong lensing regimes was not considered.\\
So, with an average of about $50$ galaxies arcmin$^{-2}$, each VLT
observation of a cluster contains about $1000$ background galaxies.\\
Now, the question that the simulations have to solve is: what is the
accuracy we can 
reach on the cosmological parameters for a given number of observed
clusters ?\\
Figure (4) gives the contours  $p_{\Omega,\Lambda}(\Omega,\Lambda)$ for 
simulations concerning $1000$ clusters in three cases : first case,
$\Omega_{o}=0.3$ and 
$\Lambda_{o}=0.7$; second case, $\Omega_{o}=0.3$ and $\Lambda_{o}=0$;
third case, $\Omega_{o}=1$ and $\Lambda_{o}=0$ . Figure 
(5) gives the same for $100$ clusters. \\
The results of these simulations are promising. They prove that with about 
$100$ clusters $\Lambda_{-0.2}^{+0.3}$ (in the case $\Omega+\Lambda=1$) and
$\Omega_{-0.25}^{+0.30}$ (in the case $\Lambda=0$) can be reached (with a
$70\%$  confidence level). $100$ clusters correspond to about $20$ nights
of VLT observation.  $\Omega=0.3$ and $\Omega=1$ universes can also be
separated at a $2\sigma=95\%$ confidence level with the same time of
observation: about 20 VLT nights. Therefore, even a modest observing
campaign on a VLT could provide interesting constraints 
 on $\Lambda$.\\
%%%%%%%%%%%%%%%%%%%%%%%%%%%%%%%%%
\begin{figure}
\begin{center}
\psfig{figure=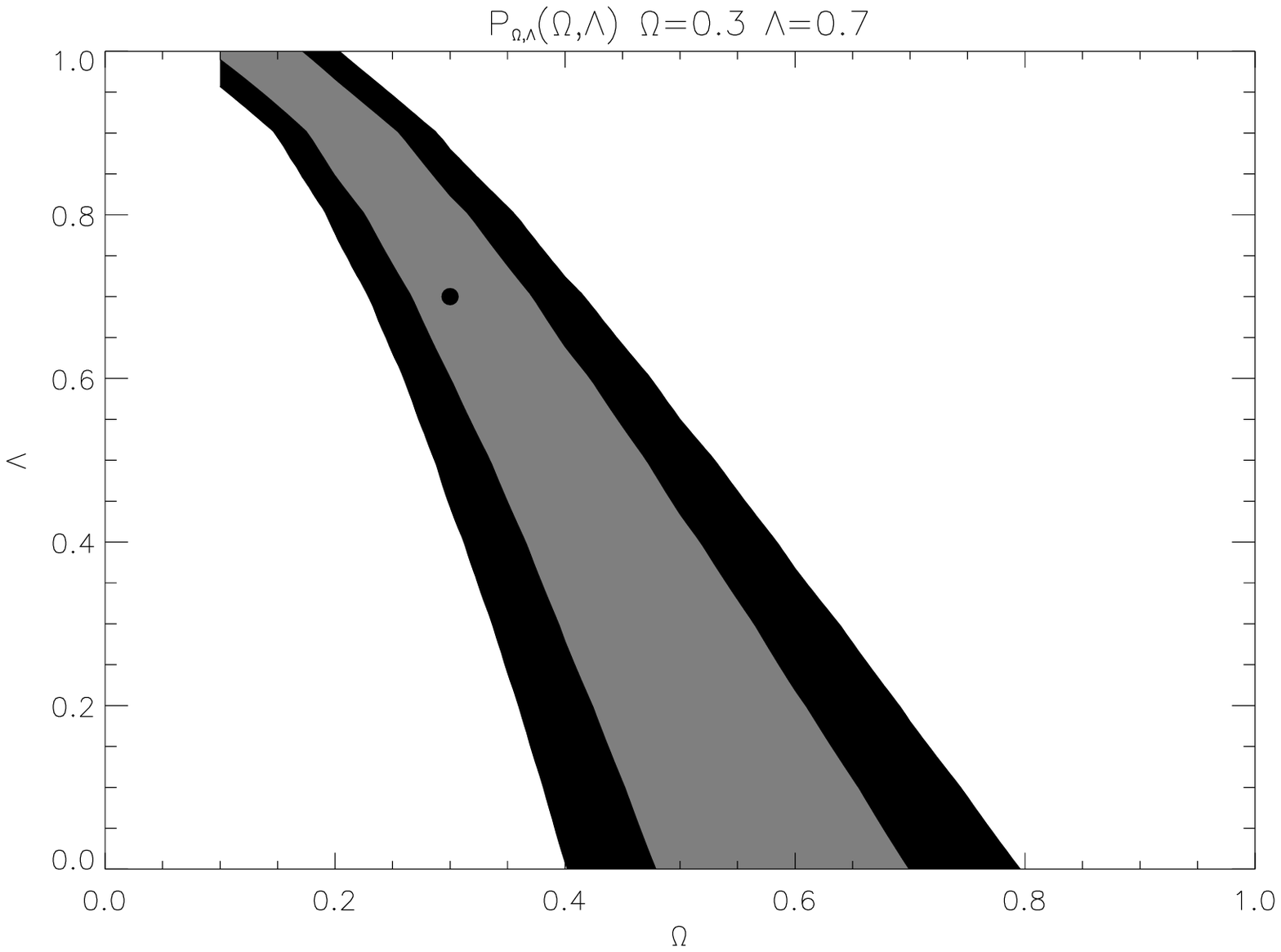,height=6cm}
\psfig{figure=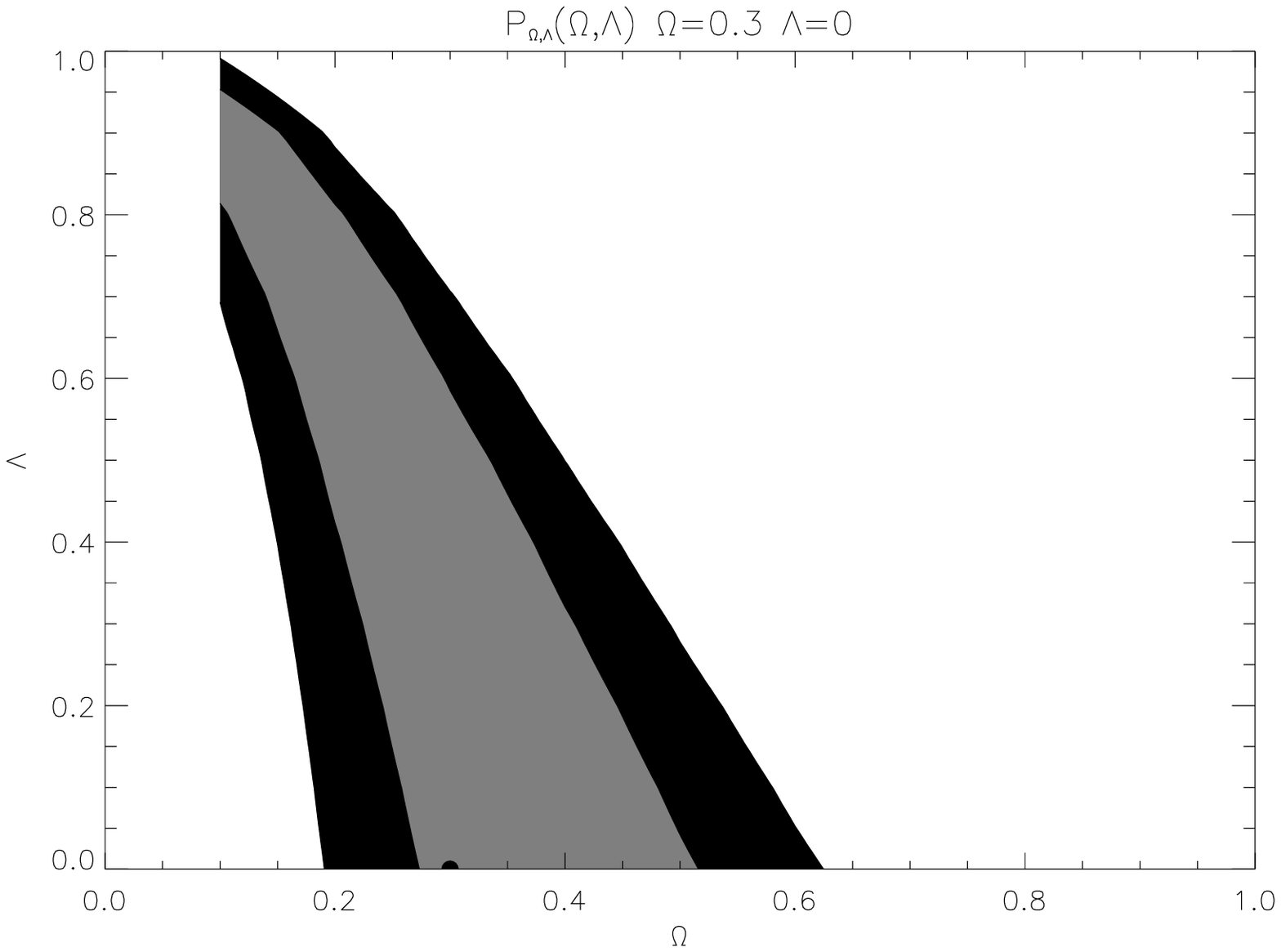,height=6cm}
\psfig{figure=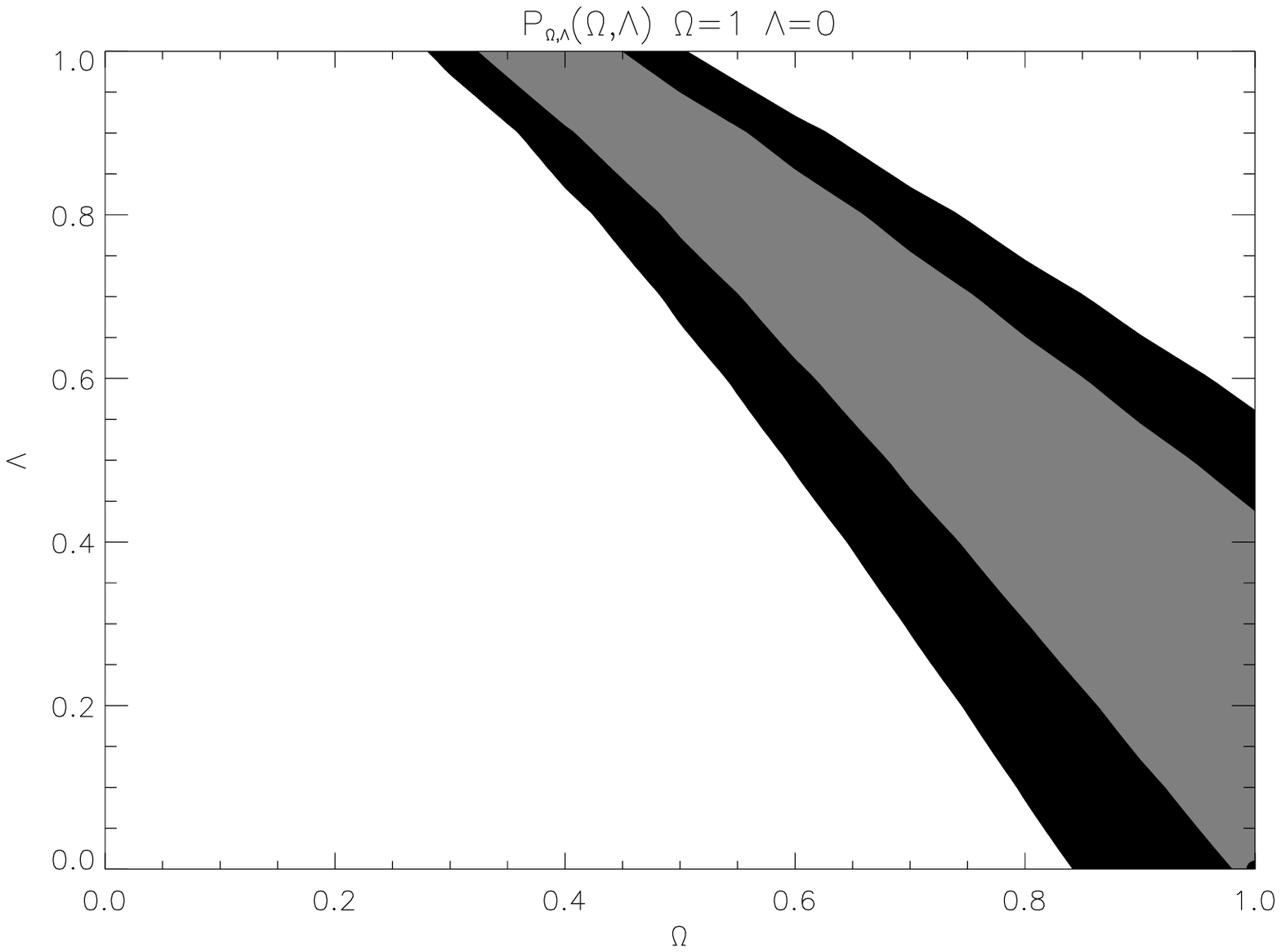,height=6cm}
\caption{Contours of the $(\Omega,\Lambda)$ probability distribution obtained 
from a simulation on $1000$ clusters in the cases
$(\Omega,\Lambda)=(0.3,0.7)$ (top 
panel) , $(\Omega,\Lambda)=(0.3,0)$ (middle
panel) and $(1,0)$ (bottom panel). We give the $1\sigma=68\%$ (grey)
 and $2\sigma=95\%$ (dark) confidence levels.}
\end{center}
\end{figure}
%%%%%%%%%%%%%%%%%%%%%%%%%%%%%%%%%%%%%
\begin{figure}
\begin{center}
\psfig{figure=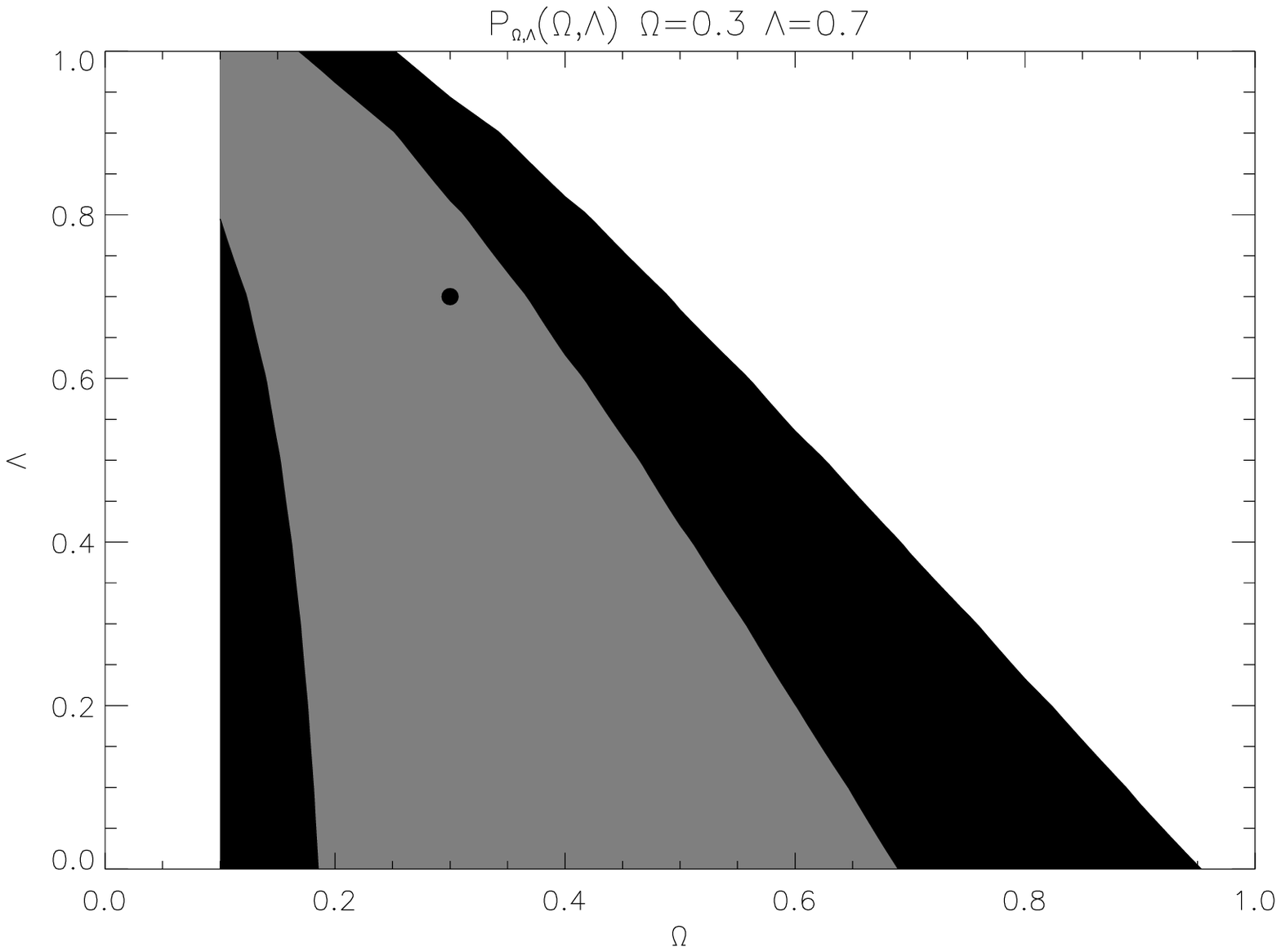,height=6cm}
\psfig{figure=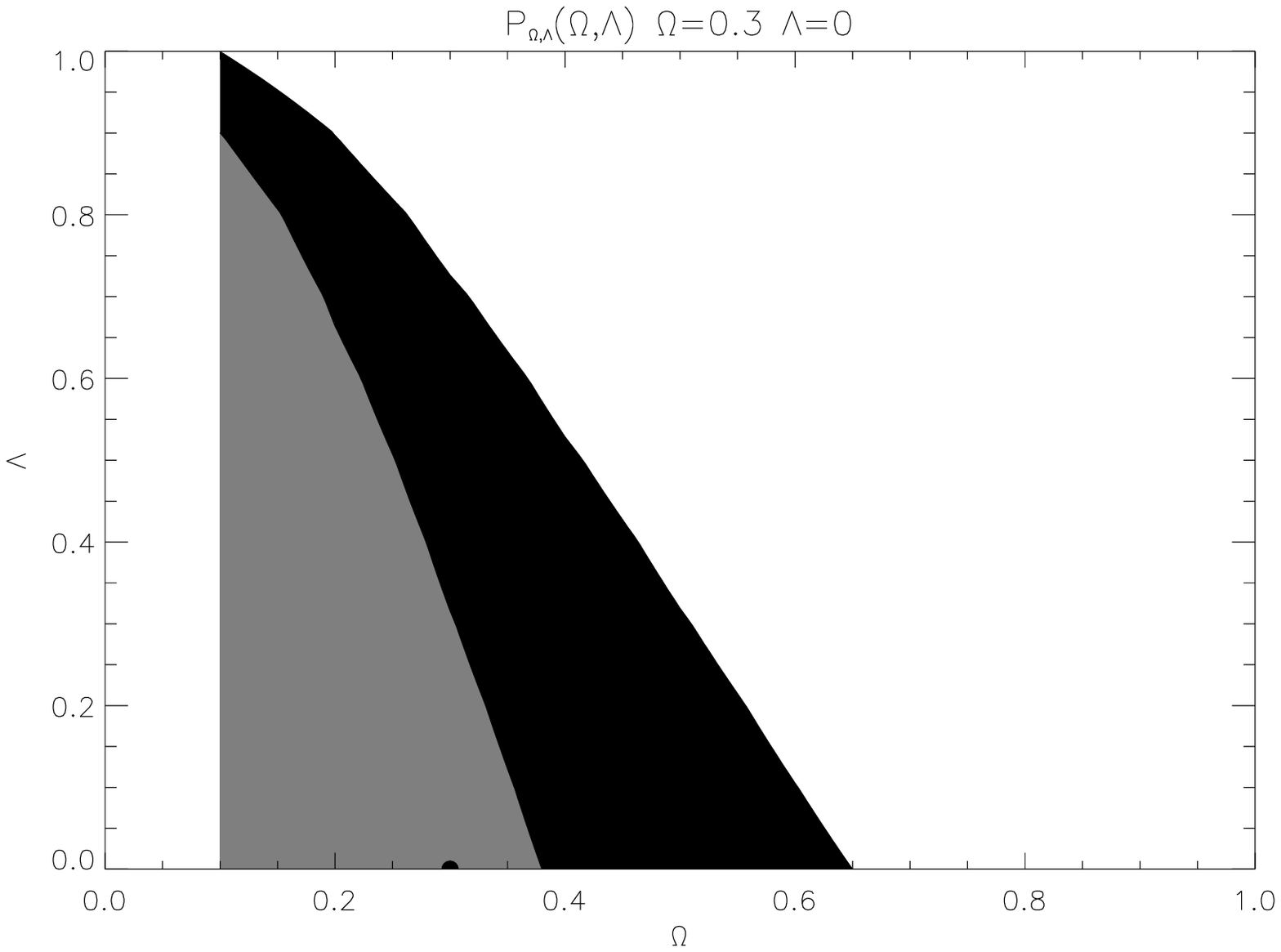,height=6cm}
\psfig{figure=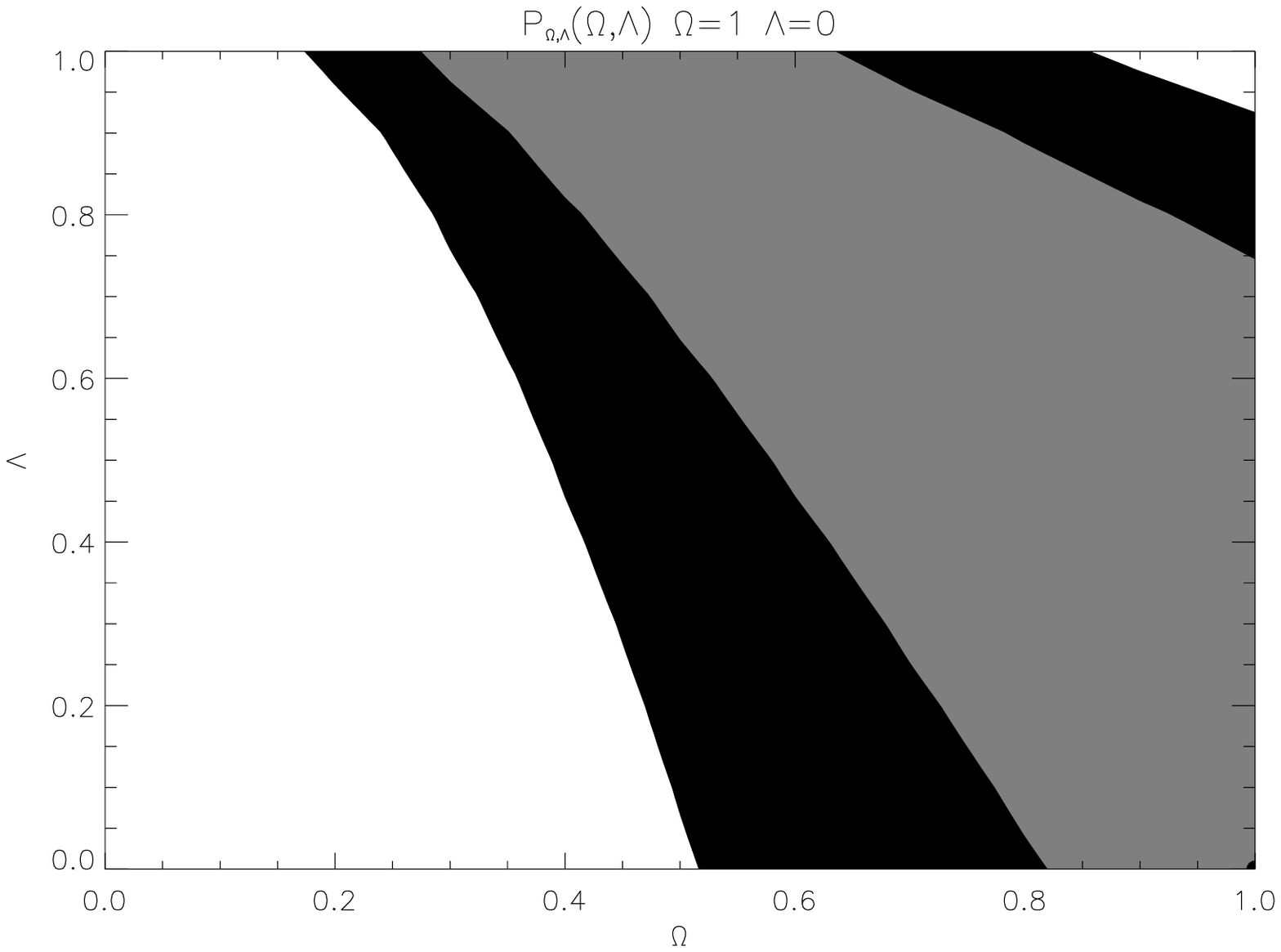,height=6cm}
\caption{Contours of the $(\Omega,\Lambda)$ probability distribution obtained 
from a simulation on $100$ clusters in the cases
$(\Omega,\Lambda)=(0.3,0.7)$ (top 
panel) , $(\Omega,\Lambda)=(0.3,0)$ (middle
panel) and $(1,0)$ (bottom panel). We give the $1\sigma=68\%$ (grey)
 and $2\sigma=95\%$ (dark) confidence levels.}
\end{center}
\end{figure}
%%%%%%%%%%%%%%%%%%%%%%%%%%%%%%%%%%%%
\section{Conclusion}
We have presented a method to constrain the cosmological parameters
$(\Omega,\Lambda)$.
  It uses gravitational distortion produced by weak lensing around clusters  
 and the photometric redshift of lensed galaxies.
 This purely geometrical method is insensitive to the lens modeling 
 and can be directly applied to real data, that is the  ellipticities of
galaxies as
observed from optical images. 
 We have calculated the statistical noise coming mainly from the intrinsic
source ellipticity
  and performed realistic simulations.\\
The main result is that with a short program of observations (about  $20$
VLT nights) 
a constraint could be provided on the value of the cosmological constant.
Using the 
observation of 100 clusters,  
we can reach $\Lambda_{-0.2}^{+0.3}$ in the case $\Omega+\Lambda=1$ or 
$\Omega_{-0.25}^{+0.3}$ in the case $\Lambda=0$ (at a $1\sigma$ level).
These results 
could be even refined down to  $10\%$ accuracy on $(\Omega,\Lambda)$ with
the use of NGST.
Indeed, the NGST looks perfectly suited for this method since it increases
the number density of observed
background galaxies. \\
We have estimated the amplitude of the systematics due to the presence of 
background structures with a multi-lensing model. It turns out that the
shift amplitudes 
on the  $(\Omega,\Lambda)$ determination are  about $0.05$.  One part of
this systematic can be directly corrected. It concerns the perturbative
effect due to galaxy-galaxy lensing (the correction of this effect need an
approximate mean potential of galaxies and will be provided by the incoming
galaxy-galaxy lensing investigations) . It also concerns the effect of
background structures integrated along the line of sight (the correction of
this effect requires calculations that takes in account the non  linear
evolution of large scale structures). It could be validated by ray tracing
simulations. \\
In conclusion, we have shown that the systematic effects could be very well
controlled by a judicious selection criteria of the clusters and the lensed
galaxies of each triplet.\\
 The $(\Omega,\Lambda)$ degeneracy of the triplet method is orthogonal to
the one of the classical $(m,z)$ of supernovae searches. In this regards,
when combined to the supernovae approach, the triplet method using  VLT or
NGST data could be extremely efficient. Therefore it seems important to
investigate more deeply the possibility to use cluster lenses as  practical
tests to constrain the curvature of the Universe.
{
\acknowledgements We thank F. Bernardeau and L. van Waerbeke for 
 fruitful discussions and their useful comments.
 This work was supported by the Programme National de Cosmologie.}
%%%%%%%%%%%%%%%%%%%%%%%%%%%%%%%%%%%%%%%%%%%%%%%%%%%%%%%%%%%%%%%%%%%%%%%%%%%%
%%%%%
% \section*{References}

%%%%%%%%%%%%%%%%%%%%%%%%%%%%%%%%%%%%%%%%%%%%%%%%%%%%%%%%%%
\section*{Appendix A : The observed ellipticity of a background source}
This appendix has two goals: firstly to recall the practical way to
calculate the image 
ellipticity $\mbox{\boldmath $\epsilon$}_{I}=\epsilon_{I} e^{2i\theta_{I}}$
of a 
source from the image of a galaxy; secondly to express this ellipticity as
a function of 
the potential of the lensing cluster through the local convergence $\kappa$
and 
complex shear $\mbox{\boldmath $\gamma$}=\gamma e^{2i\theta_{L}}$ (as it will 
be showed, the dependency in $\kappa,\gamma$ holds in the complex term 
$\mbox{\boldmath $g$}=\frac{\mbox{\boldmath $\gamma$}}{1-\kappa}$) and the 
intrinsic source ellipticity $\mbox{\boldmath $\epsilon$}_{S}=\epsilon_{S} 
e^{2i\theta_{S}}$.\\
For each background galaxy we can calculate a second order momentum matrix 
$M^{I}$, either directly from the image of the galaxy (see Bonnet 1995) 
either from the ACF of the single galaxy (see Van Waerbeke et al. 1997). Then 
the complex image ellipticity is derived from the momentum matrix through
(see 
Seitz \& Schneider 1997):

\begin{eqnarray}
   \mbox{\boldmath $\epsilon$}_{I}=\frac{M^{I}_{11}-M^{I}_{22} + 
2iM^{I}_{12}}{M^{I}_{11}+M^{I}_{22}+2\sqrt{det(M^{I})}}.
\end{eqnarray}

It is easy to check that this definition stays in agreement with equation
(5). We also 
use a second order momentum matrix $M^{S}$ for the source as it would be seen 
with no lens.\\
The effect of the lens on the ellipticity of the background galaxy can then
be 
summarized in the matrix equation (see Bonnet 1995):
\begin{eqnarray}
   M^{I} &=& \frac{AM^{S\ t}A}{|A|^{2}} \ \mbox{with} \\
   A      &=& \frac{(1-\kappa)I^{0}+\gamma J^{2\theta}}{(1-\kappa)^{2}-
\gamma^{2}}. 
\end{eqnarray}
In what follows and above, we use the notations
\begin{eqnarray}
   I^{2\theta} = \left( \begin{array}{lr}
                 {\rm cos}(2\theta) & -{\rm sin}(2\theta)                \\
                 {\rm sin}(2\theta) & {\rm cos}(2\theta) 
                                           \end{array}
                                                           \right), \\
  J^{2\theta} = \left( \begin{array}{lr}
                 {\rm cos}(2\theta) & {\rm sin}(2\theta)                \\
                 {\rm sin}(2\theta) & -{\rm cos}(2\theta) 
                                           \end{array}
                                                           \right). 
\end{eqnarray}
\noindent Equation (46) leads to 
\begin{eqnarray}
  M^{I}=\frac{(I^{0}+gJ^{2\theta})I^{\theta}
						\left( \begin{array}{lr}
                 \frac{1+\epsilon}{1-\epsilon} & 0               \\
                 0 &  \frac{1-\epsilon}{1+\epsilon}
                                           \end{array}
\right)I^{-\theta}(I^{0}+gJ^{2\theta})}{|1-g^{2}|}. 
\end{eqnarray}
\noindent Using the following properties :
\begin{eqnarray}
  J^{\alpha}J^{\beta}=I^{\alpha-\beta}, \
J^{\alpha}I^{\beta}=J^{\alpha-\beta}, \ 
I^{\alpha}J^{\beta}=J^{\alpha+\beta},
\end{eqnarray}
\noindent equation (51) gives:
\begin{equation}
\begin{array}{ll}
|1-g^{2}|(1-\epsilon_{S}^{2})\ M^{I}= &(1+\epsilon_{S}^{2})(1+g^{2})\\
\nonumber & +4g\epsilon_{S}{\rm cos}(2(\theta_{S}- \theta_{L}))]I^{0} \\
\nonumber & +2\epsilon_{S}J^{2\theta_{S}}+2g(1+\epsilon_{S}^{2}) 
J^{2\theta_{L}}\\
\nonumber & +2g^{2}\epsilon_{S}J^{4\theta_{L}-2\theta_{S}}.   
\end{array}
\end{equation}
Finally, whatever the value of $g$ is (meaningly either in the weak or
strong lensing 
regime) the measured image ellipticity writes:
\begin{eqnarray}
  \mbox{\boldmath $\epsilon$}_{I}=\frac{\mbox{\boldmath 
$\epsilon$}_{S}+(1+\epsilon_{S}^{2})\mbox{\boldmath $g$}+\mbox{\boldmath 
$\epsilon$}_{S}^{*}\mbox{\boldmath $g$}^{2} 
}{max(1,g^{2})+\epsilon_{S}^{2}min(1,g^{2})+ \mbox{\boldmath 
$\epsilon$}_{S}^{*}\mbox{\boldmath $g$}+ \mbox{\boldmath 
$\epsilon$}_{S}\mbox{\boldmath $g$}^{*} },
\end{eqnarray}
\noindent which, in the weak lensing and arclet regimes ($g<1$) simplifies into (to
the third 
order in $\epsilon_{S}$ and $g$):
\begin{eqnarray}
  \mbox{\boldmath $\epsilon$}_{I}=(1-g^{2})\mbox{\boldmath $\epsilon$}_{S} + 
\mbox{\boldmath $g$}-\mbox{\boldmath $g$}^{*}\mbox{\boldmath 
$\epsilon$}_{S}^{2} ,
\end{eqnarray}
\noindent where the term $\mbox{\boldmath $g$}^{*}\mbox{\boldmath $\epsilon$}_{S}^{2}$ 
is negligible regarding to the term $\mbox{\boldmath $\epsilon$}_{S}$ and
vanishes 
in $1/\sqrt{N}$ (because the argument of this term behaves randomly) in the 
operator G as the different noises described in section 3.4. Therefore the
final 
equation for the image ellipticity writes :
\begin{eqnarray}
  \mbox{\boldmath $\epsilon$}_{I}=(1-g^{2})\mbox{\boldmath $\epsilon$}_{S} + 
\mbox{\boldmath $g$}.
\end{eqnarray}

It is worht noticing  here that the term $(1-g^{2})$ is in favor of 
the method developed in this paper. Indeed, with the conditions used in the 
simulations (section 6), the mean value of this term is about $0.85$. It
means that 
the noise coming from the intrinsic source ellipticity is lowered by $15\%$ 
or, for a given signal to noise ratio, the number of required clusters is
lowered by 
$28\%$.

\section*{Appendix B : The case of a perturbing lens}
This section will study the influence of a perturbative lens (galaxy,
cluster or higher 
scale structure) located behind the principal lensing cluster on the measured 
ellipticity of a background source.\\
The calculation of the total amplification matrix in the case of two (or
more) consecutive lenses 
has already been done (see for example Kovner 1997). The result 
can be simply noticed as follows:

\begin{eqnarray}
  A^{-1} &=& I^{0}-L-L^{P}+cL^{P}L \ \mbox{with} \\
  L        &=&  \kappa I^{O}+\gamma J^{2\theta_{L}} \ \mbox{and} \\
  L^{P}  &=&  \kappa^{P} I^{O}+\gamma^{P} J^{2\theta_{P}}.
\end{eqnarray}

The $^{P}$ upper index refers to the perturbative lens; $\kappa^{P}$ is
defined as 
$\kappa$ in equation (3), just changing the principal lens noticed $l$ by the 
perturbative lens $l^{P}$ for the definition of the perturbative critical
surface mass 
density $\Sigma_{crit}^{P}$. The $c$ coupling factor is 
$c=\frac{D_{LL^{P}}D_{OS}}{D_{OL^{P}}D_{LS}}$.\\
Equation (57) rewrites:

\begin{eqnarray}
  A^{-1} =&
 (1-\kappa-\kappa^{P}+c\kappa \kappa^{P})I^{0}-\gamma(1-
c\kappa^{P})J^{2\theta_{L}} \\
\nonumber &-\gamma^{P}(1-c\kappa)J^{2\theta_{P}}+ c\gamma 
\gamma^{P} I^{2(\theta_{P}-\theta_{L})},
\end{eqnarray}  

which can be inverted into

\begin{eqnarray}
  A =\frac{1}{|A^{-1}|}&
[(1-\kappa-\kappa^{P}+c\kappa \kappa^{P})I^{0}+\gamma(1-
c\kappa^{P})J^{2\theta_{L}}  \\
\nonumber &+\gamma^{P}(1-c\kappa)J^{2\theta_{P}}+ c\gamma 
\gamma^{P} I^{2(\theta_{L}-\theta_{P})}] .
\end{eqnarray}  

Contrary to the case of a single lens where the Amplification matrix is
symmetrical, 
here appears an anti-symmetric term, the last one of the above equation.
However, 
even if the Amplification matrix is non symmetric, it keeps the propriety
to transform 
an ellipse into another ellipse from the source plan to the image plan.
Indeed the 
transformation from the source to the image can be represented by the vector 
notation $X_{S}=A^{-1} X_{I}$. To say that the source is an ellipse is
equivalent to say 
that exists a matrix $E$ such that $^{t}X_{S}^{t}EEX_{S}=1$ (it comes from
the fact 
that every positive symmetric matrix can be written as $^{t}EE$).  So the
position of 
image vectors writes   $^{t}X_{I}^{t}\tilde{E}\tilde{E}X_{I}=1$ with
$\tilde{E}=EA^{-1}$, 
which proves that the image is another ellipse. We thus can search for the
image 
ellipticity of a source distorted by both lenses, as a function of the
source ellipticity 
and the potential of the principal and perturbative lenses. We proceed the
same way 
as in annex A.\\
In the following, we only keep second order terms in 
$\kappa,\kappa^{P},\gamma,\gamma^{P}\ and \ \epsilon_{S}$. So we can use the 
definitions: 

\begin{eqnarray}
  \tilde{\mbox{\boldmath $g$}} &=& (1-\kappa-(1-c)\kappa^{P})^{-
1}\mbox{\boldmath $\gamma$} \\
 \tilde{\mbox{\boldmath $g$}}^{P} &=& (1-(1-c)\kappa-\kappa^{P})^{-1} 
\mbox{\boldmath $\gamma$}^{P} \ ,
\end{eqnarray}
%%%%%%%%%%%%%%%%%%%%%%%%%%%%%%%%%%%
and rewrite the amplification matrix:
%%%%%%%%%%%%%%%%%%%%%%%%%%%%%%%%
\begin{eqnarray}
  A &=& \frac{I^{0}+\tilde{g}J^{2\theta_{L}}+\tilde{g}^{P}J^{2\theta_{P}}+ 
c\tilde{g}\tilde{g}^{P}
I^{2(\theta_{L}-\theta_{P})}}{1-\tilde{g}^{2}-\tilde{g}^{P2}-2(1-
c)Re(\tilde{g}\tilde{g}^{P*})}.
\end{eqnarray}
%%%%%%%%%%%%%%%%%%%%%%%%%%%%%%%%%
Finally, with a calculation similar to the one done in annex A, we obtain
for the 
image ellipticity:
%%%%%%%%%%%%%%%%%%%%%%%%%%%%%%%%
\begin{eqnarray}
  \mbox{\boldmath $\epsilon$}_{I}=
&(1-|\tilde{\mbox{\boldmath 
$g$}}+\tilde{\mbox{\boldmath $g$}}^{P}|^{2})\mbox{\boldmath $\epsilon$}_{S} + 
\tilde{\mbox{\boldmath $g$}}+\tilde{\mbox{\boldmath $g$}}^{P} \\
\nonumber &-
4cRe(\tilde{\mbox{\boldmath $g$}}\tilde{\mbox{\boldmath 
$g$}}^{P*}(\tilde{\mbox{\boldmath $g$}}+\tilde{\mbox{\boldmath $g$}}^{P})-
2c\tilde{\mbox{\boldmath $g$}}^{*}\tilde{\mbox{\boldmath
$g$}}^{P}\mbox{\boldmath 
$\epsilon$}_{S}-\epsilon_{S}^{2}(\tilde{\mbox{\boldmath 
$g$}}^{*}+\tilde{\mbox{\boldmath $g$}}^{P*})\,
\end{eqnarray}
%%%%%%%%%%%%%%%%%%%%%%%%%%%
which we simplify into
%%%%%%%%%%%%%%%%%%%%%%
\begin{eqnarray}
  \mbox{\boldmath $\epsilon$}_{I}=(1-|\tilde{\mbox{\boldmath 
$g$}}+\tilde{\mbox{\boldmath $g$}}^{P}|^{2})\mbox{\boldmath $\epsilon$}_{S} + 
\tilde{\mbox{\boldmath $g$}}+\tilde{\mbox{\boldmath $g$}}^{P}.
\end{eqnarray}

In this last equation we have cancelled all the terms higher to the second
order and 
oriented randomly (we consider that the source ellipticity, the principal and 
perturbative shears are independent. Terms higher to the second order with a 
random orientation give a negligible noise -regarding to the noise coming
from the 
intrinsic source ellipticity- in the calculation of $G$).\\
Comparing equations (62) and (66) we can see the effect of the perturbative
lens on 
the measured ellipticity:

\begin{enumerate}
 \item  Firstly a complex additive correction $\tilde{\mbox{\boldmath
$g$}}^{P}$ which orientation is 
non correlated to the one of the cosmological term.
 \item  Secondly a scalar correction $(1-c)\kappa^{P}$ on the 
cosmological term $\mbox{\boldmath $g$}$, due to coupling between the main
and the perturbative lenses. \\
\end{enumerate}

The simple form of equation (66) can be easily generalized to the case of a
succession 
of many lenses noticed $L^{0}$ (instead of the principal lens), $L^{1}$
(instead of the 
perturbative lens), $L^{2}$, ..., $L^{n}$ between the observer and the
source noticed 
$S$. If the coupling factor between the lenses $i$ and $j$ (with
$z_{i}<z^{j}$) is noticed 
$c_{i,j}^{S}=\frac{D_{OS}D_{L^{i}L^{j}}}{D_{OL_{j}}D_{L_{i}S}}$, then
equation 
(55) giving the amplification matrix can be generalized into:

\begin{eqnarray}
  A^{-1} = \bigotimes_{i=n}^{i=0}(I^{0}-L^{i}) =
(I^{0}-L^{n})\otimes...\otimes (I^{0}-L^{0}) , 
\end{eqnarray}

where the $\otimes$ symbol defines a particular matrix product with the
following properties: $I^{0}\otimes L^{i}=L^{i}\otimes I^{0}=L^{i}$,
$L^{j}\otimes L^{i}=c_{i,j}^{S}L^{j}L^{i}$, $L^{k}\otimes L^{j}\otimes
L^{i}=c_{i,j}^{S}c_{j,k}^{S}L^{k}L^{j}L^{i}$ etc...\\

We can remark that in the case of two very close lenses, their associated
coupling 
factor $c$ is zero. With the same calculations and approximations as above we 
obtain:

\begin{eqnarray}
  \mbox{\boldmath $\epsilon$}_{I} &=&
(1-|\sum_{i=0}^{i=n}\tilde{\mbox{\boldmath $g$}}^{i}|^{2}) \mbox{\boldmath 
$\epsilon$}_{S}+\sum_{i=0}^{i=n}\tilde{\mbox{\boldmath $g$}}^{i} \ {\rm with} \\
  \tilde{\mbox{\boldmath $g$}}^{i} &=& (1-\kappa^{i}-\sum_{j\neq i}
(1-c_{i,j}^{S})\kappa^{j})^{-1} \gamma^{i}.
\end{eqnarray}

\end{document}